\documentclass[aps,twocolumn,prl,floatfix,superscriptaddress,showpacs]{revtex4-1}

\usepackage{epsf}
\usepackage{epsfig}
\usepackage{graphicx}
\usepackage{dcolumn}
\usepackage{braket}
\usepackage{bm}
\usepackage{amsfonts}
\usepackage{amsmath}
\usepackage{amssymb}
\usepackage{color,soul}
\usepackage{natbib}

\definecolor{light-gray}{gray}{0.5}

\newcommand{\sgn}{\mathrm{sgn}}

\begin{document}

\title{Floquet interface states in illuminated three-dimensional topological insulators}

\author{H.~L.~Calvo}
\affiliation{Instituto de F\'{\i}sica Enrique Gaviola (CONICET) and FaMAF, Universidad Nacional de C\'ordoba, Argentina}

\author{L.~E.~F.~Foa Torres}
\affiliation{Instituto de F\'{\i}sica Enrique Gaviola (CONICET) and FaMAF, Universidad Nacional de C\'ordoba, Argentina}

\author{P.~M.~Perez-Piskunow}
\affiliation{Instituto de F\'{\i}sica Enrique Gaviola (CONICET) and FaMAF, Universidad Nacional de C\'ordoba, Argentina}

\author{C.~A.~Balseiro}
\affiliation{Centro At\'omico Bariloche and Instituto Balseiro, Comisi\'on Nacional de Energ\'ia At\'omica, 8400 Bariloche, Argentina}
\affiliation{Consejo Nacional de Investigaciones Cient\'ificas y T\'ecnicas (CONICET), Argentina}

\author{Gonzalo~Usaj}
\affiliation{Centro At\'omico Bariloche and Instituto Balseiro, Comisi\'on Nacional de Energ\'ia At\'omica, 8400 Bariloche, Argentina}
\affiliation{Consejo Nacional de Investigaciones Cient\'ificas y T\'ecnicas (CONICET), Argentina}

\begin{abstract}
Recent experiments showed that the surface of a three dimensional topological insulator develops gaps in the Floquet-Bloch band spectrum when illuminated with a circularly polarized laser. These Floquet-Bloch bands are characterized by non-trivial Chern numbers which only depend on the helicity of the polarization of the radiation field. Here we propose a setup consisting of a pair of counter-rotating lasers, and show that one-dimensional chiral states emerge at the interface between the two lasers. These interface states turn out to be spin-polarized and may trigger interesting applications in the field of optoelectronics and spintronics.
\end{abstract}



\maketitle

\textit{Introduction.--} 
Amid the thrill sparked by graphene~\cite{Novoselov2005a,Zhang2005} and its record properties~\cite{Geim2007}, the discovery of topological insulators (TIs)~\cite{Koenig2007,Hsieh2008} developed with surprising speed. Indeed, TIs were predicted two years earlier in graphene~\cite{Kane2005}, but the necessary spin-orbit interactions were too weak for this to be observed and a different playground was needed to realize them~\cite{Bernevig2006}. Most TIs are three-dimensional materials like usual solids, but with a special property: they have a bulk band gap while keeping states that propagate with unprecedented robustness at the periphery of the sample~\cite{Fu2007,Hasan2010}. These peculiar states hold great promise for quantum computation~\cite{Moore2010} but at the same time open up a major challenge: controlling them is particularly demanding for 3D TIs.

Encompassing the rapid progress in graphene photonics~\cite{Bonaccorso2010} and optoelectronics~\cite{Glazov2014,Hartmann2013}, theoretical studies predicted the formation of laser-induced band gaps~\cite{Oka2009} in graphene when properly tuning the laser polarization, frequency and intensity~\cite{Calvo2011,Zhou2011,Savelev2011,SuarezMorell2012}. More recently, these gaps were unveiled at the surface of a TI through ARPES~\cite{Wang2013}. This triggered great expectations for achieving laser-assisted control not only in the form of an on-off switch for the available states but also because theoretically non-trivial topological states~\cite{Oka2009,Lindner2011,Kitagawa2011} can be induced on a diversity of materials~\cite{Piskunow2014,Sentef2014,Dahlhaus2014,Quelle2014,Lopez2015}, and also in cold matter physics~\cite{Goldman2014,Choudhury2014}. Exciting questions arise about the nature of these novel states~\cite{Gomez-Leon2014, Bilitewski2015, TenenbaumKatan2013a, Rudner2013, Ho2014, Zhou2014, Usaj2014, DAlessio2014, Dehghani2014, Liu2015, Seetharam2015, Iadecola2015, Gu2011, Kundu2014, Foa2014, Dehghani2015}, the possibilities for manipulating them~\cite{TenenbaumKatan2013a}, the associated topological invariants~\cite{Rudner2013,Ho2014,Zhou2014,Usaj2014,DAlessio2014}, their statistical properties~\cite{Dehghani2014, Liu2015, Seetharam2015, Iadecola2015} and their two-terminal~\cite{Gu2011,Kundu2014} and multi-terminal (Hall) response~\cite{Foa2014,Dehghani2015}. Still, an experimental realization of the Floquet chiral edge states is missing. Most studies considered two-dimensional systems, except for Refs.~\cite{Lindner2013} and~\cite{Tenenbaum-Katan2013} where the target was a 3D semiconductor.

\begin{figure}[tb]
 \includegraphics[width=0.9\columnwidth]{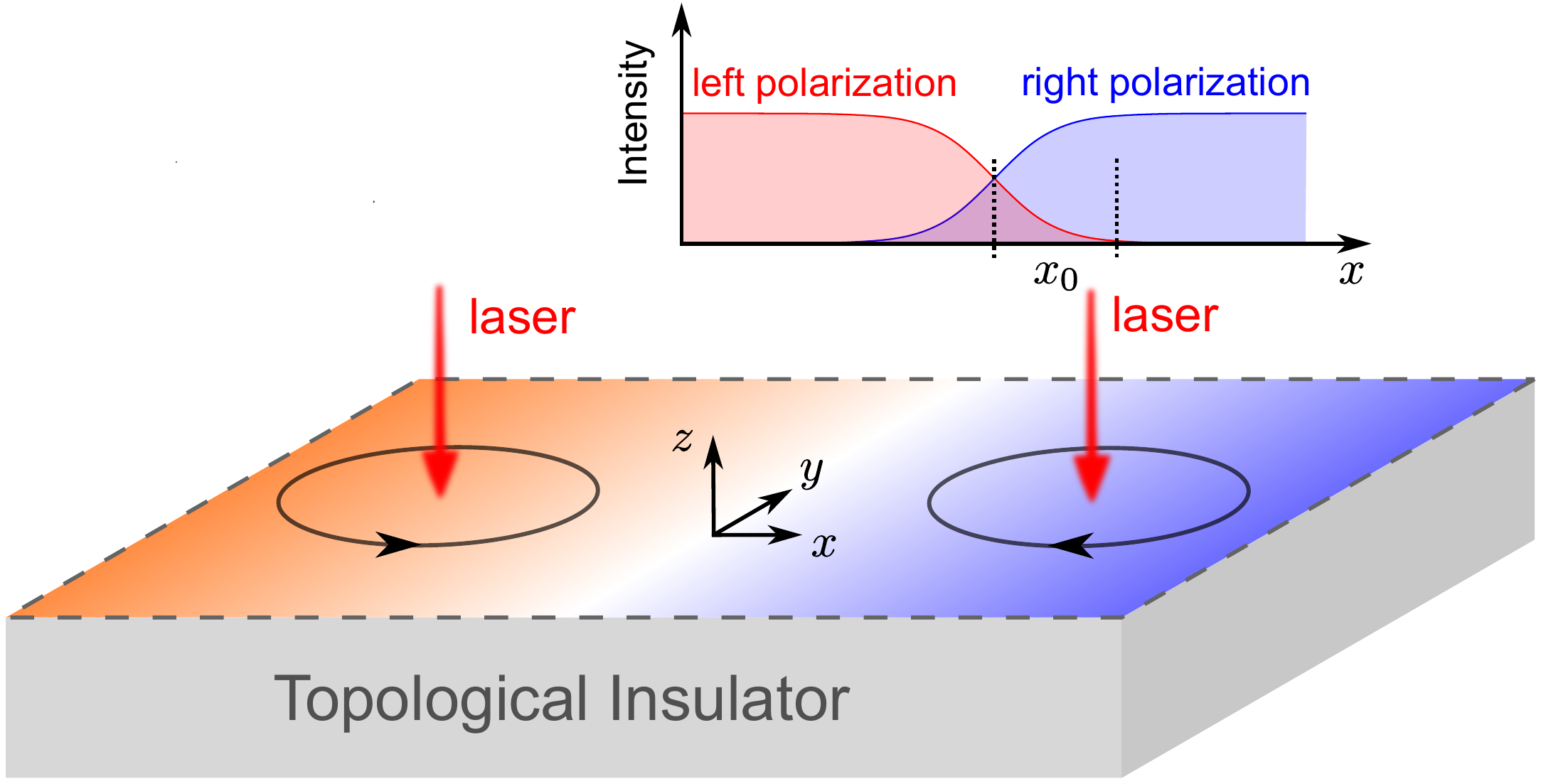}
 \caption{(color online). Scheme of the proposed setup where the surface of a 3D TI is illuminated by two circularly polarized lasers with opposite directions of rotation. The $y$-axis defines the interface region between the two lasers. The dashed line in the border indicates that the system extends indefinitely in the $x$-$y$ plane. Inset: Laser intensities as function of the $x$-coordinate. At the interface, the two lasers interfere and the total field becomes linearly polarized.}
 \label{fig:1}
\end{figure}

Here we show that besides opening a band gap as in Ref.~\cite{Wang2013}, illuminating a 3D TI with a suitable set of lasers can confine the surface states into one-dimensional states which also bear a topological origin. The proposed setup is represented in Fig.~\ref{fig:1}: Two lasers with opposite circular polarization incident perpendicularly to a face of a 3D TI. This can be devised through, \textit{e.g.}, a single laser beam splitted into two with opposite helicity. The interface between the lasers is assumed to be shorter than the thermalization length so that the occupations are determined by the (larger) regions without radiation. As we will see below, this modification of the experimental setup in Ref.~\cite{Wang2013} introduces Floquet states propagating along the boundary where the polarization changes. Our results follow from simulations of the Floquet spectra based on low-energy models, which are further supported by: (\textit{i}) a calculation of the topological invariants and (\textit{ii}) explicit calculations for a driven 3D lattice model. Interestingly, we show that the resulting Floquet boundary states, which arise from a topological transition between the illuminated regions, carry spin-polarized currents.

\textit{Illuminated TIs and Floquet theory.--} We consider a low-energy Hamiltonian describing the surface of a TI. Assuming the $(00\bar{1})$ direction~\cite{Zhang2012} and linear order in $\bm{k}$, the effective surface Hamiltonian reads $H_0 = \hbar v(k_y \sigma_x - k_x \sigma_y)$, where $\sigma_x$ and $\sigma_y$ are Pauli matrices describing the spin degree of freedom. The time-dependent field is included through the Peierls substitution $\bm{k} \rightarrow \bm{k}+e\bm{A}(t)/\hbar c$, with $\bm{A}(t)$ the laser's vector potential. In the regions dominated by one of the two lasers, \textit{i.e.} $|x|\gg x_0$, with $x_0$ the characteristic length of the lasers' interface (see Fig.~\ref{fig:1}), we choose a circularly polarized field $\bm{A}_\tau(t) = A_0 [\cos(\tau \Omega t + \varphi)\, \bm{e}_x + \sin(\tau \Omega t + \varphi)\, \bm{e}_y]$, where $\tau = \pm 1$ sets the direction of rotation and $\varphi$ determines its orientation (measured from the $x$-axis) at $t=0$ and, as shown later on, it becomes relevant at the interface's region $x \sim 0$. The time-dependent Hamiltonian thus reads 
\begin{equation}
H_\tau(t) = H_0 + \gamma \sin(\tau \Omega t + \varphi)\, \sigma_x - \gamma \cos(\tau \Omega t + \varphi)\, \sigma_y,
\end{equation}
where $\gamma = v e A_0/c$ characterizes the strength of the perturbation. A suitable description of the dc-spectrum and the topological properties 
of the system can be achieved through the  Floquet theory. By using Floquet's theorem, we obtain a time-independent Hamiltonian 
in Floquet space, defined as the direct product $\mathcal{R} \otimes \mathcal{T}$ between the usual Hilbert space $\mathcal{R}$ and the space 
of time-periodic functions $\mathcal{T}$. This space is spanned by the states $\ket{\Psi_\sigma,m}$, where $\sigma = \{\uparrow,\downarrow \}$ 
accounts for spin and $m$ is the Fourier index. The Floquet Hamiltonian writes
\begin{equation}
\mathcal{H}_F^\tau(\bm{k}) = H_0 \otimes I + I \otimes N_\Omega  + i \gamma \tau \sum_{\beta=\pm} \beta\, e^{-i \beta \tau \varphi} \sigma_{\beta\tau} \otimes \Delta_{\beta}, 
\label{eq:Floquet}
\end{equation}
where we use $\sigma_\pm = (\sigma_x\pm i \sigma_y)/2$. Such Hamiltonian can be imagined as a series of replicas (Floquet channels) of $H_0$, each 
one defined in a Fourier component of the driving. The static $H_0$ enters in the diagonal part, together with the contribution $[N_\Omega]_{n,m} = m \hbar \Omega \delta_{n,m}$ from the driving field, and the vector potential couples, through $[\Delta_{\beta}]_{n,m} = \delta_{m,n-\beta}$, those channels differing in their Fourier indices by $\Delta m = \pm 1$.

For the calculation of the laser-induced band gaps and the associated Chern numbers, it is enough to consider an homogeneous system defined at the TI's surface through Eq.~(\ref{eq:Floquet}). The underlying assumption is that $\hbar \Omega$ is smaller than the bulk gap, such that the states associated to the bulk do not participate in the gap openings. As discussed in Refs.~\cite{Piskunow2014,Usaj2014}, these laser-induced gaps are indeed depletions of the time-averaged density of states which results from weighting the Floquet spectrum on the $m=0$ channel. By assuming low intensities ($\gamma/\hbar\Omega \ll 1$) we restrict ourselves to the main contributions to the band gap openings around $\varepsilon = \hbar\Omega/2$ and at the Dirac point $\varepsilon = 0$, henceforth called the zone boundary (ZB) and the zone center (ZC) gaps, respectively. These two gaps were described in Ref.~\cite{Fregoso2013}, obtaining $\Delta_1 \approx \gamma$ and $\Delta_0 \approx 2\gamma^2/\hbar\Omega$ for the ZB and ZC gaps, respectively. Notice that a $\pi/2$-rotation along the $z$-direction of the spin coordinate system maps $H_0$ to the low-energy Hamiltonian describing a single valley in graphene ($H_0 \rightarrow \hbar v\bm{k}\cdot \boldsymbol{\sigma}$). Therefore, apart from a change in the Fermi velocity, the laser-induced gaps show the same dependencies in both systems~\cite{Syzranov2008,Oka2009,Calvo2011}. 

The equivalence between Eq.~(\ref{eq:Floquet}) and the low-energy description for illuminated graphene 
can be exploited even further: In graphene, the laser-induced gaps are characterized by non-trivial Chern numbers, and the bulk-boundary 
correspondence leads to Floquet chiral states at the sample edges~\cite{Gu2011,Piskunow2014,Usaj2014}. Can similar states appear here? A first problem is simply that the surface of a 3D solid cannot have a boundary. This motivates our proposal of changing the light polarization as in Fig. \ref{fig:1}, thereby introducing an effective boundary (much like a domain wall, as discussed in other examples~\cite{Fu2008,Chu2011,Zhang2013}) where Floquet chiral states develop---by chiral we mean that the direction of motion is fixed by the helicity of the two lasers.

Starting from Eq.~(\ref{eq:Floquet}), we proceed as in Refs.~\cite{Usaj2014,Piskunow2015}: First we isolate each crossing where a 
band gap opens, and then we compute a $2 \times 2$ effective Hamiltonian of the form $\mathcal{H}_\mathrm{eff} = \bm{h}\cdot\bm{\sigma}$. 
The contribution to the Chern number of the lower band can be calculated through the expression~\cite{Hasan2010}:
\begin{equation}
C = \frac{1}{4\pi} \int \mathrm{d}^2\bm{k}\, \frac{\bm{h}}{h^3}\cdot\left( \partial_{k_x}\bm{h} \times \partial_{k_y}\bm{h} \right).
\end{equation}
In the ZB gap region, the band gap opening comes from the crossing between the states $\ket{\Psi_{+},0}$ and $\ket{\Psi_{-},1}$. Here $\ket{\Psi_{\pm}}$ refer to the conduction and valence band solutions of $H_0$, respectively, and the second index ($0$ or $1$) indicates the Floquet channel. By reducing the Floquet Hamiltonian to these states, we obtain that the contribution to the Chern number is $C_1=\tau$. In the ZC region there are two processes related to the gap opening. These consist of (\textit{i}) the renormalization of $\ket{\Psi_{\pm},0}$ due to the coupling to the $m=\pm 1$ states, and (\textit{ii}) the level crossing between $\ket{\Psi_{-},1}$ and $\ket{\Psi_{+},-1}$, bridged by the $m=0$ states. A straightforward calculation of these two contributions yields $C_0=-\tau/2+2\tau = 3\tau/2$. While in graphene this half-integer Chern number is compensated by spin and valley degeneracies, in strong TIs, where the surface encloses an odd number of Kramers degenerate Dirac points, a half-integer Chern number results for example when exposing the material to a static magnetic field~\cite{Hasan2010,Niemi1983,Redlich1984,Qi2008,Chu2011,Zhang2015}.


\textit{Interface states in 3D TIs.--} A natural question relies on the bulk-boundary correspondence in illuminated TIs associated to the non-zero Chern numbers of the Floquet bands. In the present case, since inverting the helicity of the circularly polarized laser changes the sign of the Chern numbers, one expects the generation of chiral states at the boundary between the two regions. 

\begin{figure}
\includegraphics[width=0.95\columnwidth]{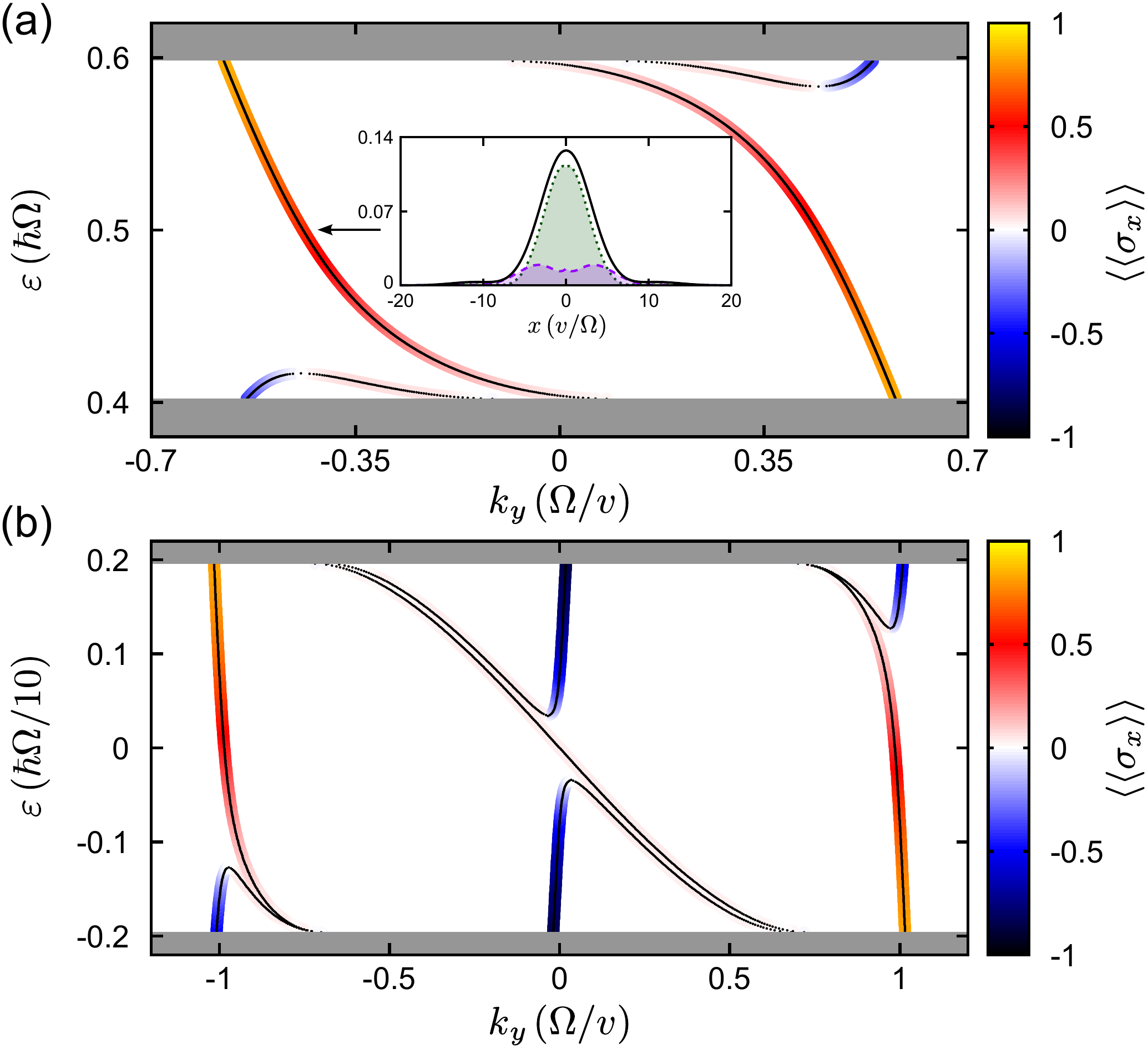}
\caption{(color online). Laser generation of interface states crossing the ZB (a) and ZC (b) gaps. Gray regions indicate extended states zones. The color scale shows the time-average spin texture $\braket{\braket{\sigma_x}}$. Inset: Time-averaged probability density $\bar{P}$ for the state at $\varepsilon = \hbar \Omega/2$ in solid line (black). Dotted line (green) shows the $m=0$ component while dashed line (purple) is for $m=1$. Here we use $\gamma / \hbar \Omega = 0.2$ and $\varphi = \pi/2$.}
\label{fig:2}
\end{figure}

To elucidate this question we proceed by solving the proposed model at the laser's interface. For simplicity in the calculation we assume a sudden change of the laser's direction of rotation by assigning a different $\tau$ to each portion of the system [according to Fig.~\ref{fig:1}, $\tau(x) \equiv -\sgn(x)$]. The resulting differential equation therefore reads
\begin{equation}
\partial_x \boldsymbol{\Psi}(\bm{r}) = \mathcal{M}_{\tau(x)} \boldsymbol{\Psi}(\bm{r}),
\label{eq:diffeq} 
\end{equation}
where $\mathcal{M}_\tau = i \sigma_y (\mathcal{H}_F^\tau(k_y\,\bm{e}_y)-\varepsilon\mathcal{I})/v$ and $\boldsymbol{\Psi}(\bm{r}) = e^{i k_y y}(\psi_{\uparrow,0}, \psi_{\downarrow,0}, \psi_{\uparrow,1}, \psi_{\downarrow,1})^\mathrm{T}$ contains the wave-function coefficients $\psi_{\sigma,m}(x)$ for the involved channels~\cite{note1}. The solutions to the above differential equation follow a standard diagonalization of $\mathcal{M}_\tau$ in the two regions with the appropriate boundary condition~\cite{Zhang2012} and is discussed in detail in the Supplemental Material~\cite{note2}.

In Fig.~\ref{fig:2} we show the resulting localized states in both the ZB and ZC gaps in a configuration of $\varphi = \pi/2$, such that at $t = 0$ the vector potentials point parallel to the interface's direction. Our calculations allow to verify the bulk-boundary correspondence. Indeed, the difference between forward and backward propagating states (relative to the $y$-axis) is always fixed to the calculated difference between the Chern invariants at each side of the interface, yielding $\Delta C_1 = -2$ for the ZB gap and $\Delta C_0 = -3$ for the ZC gap. Other choices in the orientation $\varphi$ lead to changes in the dispersions of the chiral states but keeping these numbers the same~\cite{note2}. Although never crossing the gaps, additional states localized at the interface are also present---since they are not chiral, one could expect these states to backscatter when encountering impurities in the sample. In Fig.~\ref{fig:2}(a) (inset) we show the time-averaged probability density, $\bar{P}(x) = \sum_{m,\sigma} |\psi_{\sigma,m}(x)|^2$, associated to one of the two localized states crossing the gap at $\varepsilon = \hbar\Omega/2$ (negative $k_y$). Here, the decay length depends inversely on the size of the gap and, similarly to illuminated graphene~\cite{Piskunow2014,Usaj2014}, it is independent of the microscopic details of the sample. In a more realistic situation where the inversion of the laser helicity is taken gradually over a finite length $x_0$ (see below), the width of these states shows to depend also on the latter parameter $x_0$. It can be seen also that there is a pronounced asymmetry in the contributions from the $m=0$ and $m=1$ channels to the overall probability density, which is particular to the relative angle between $\varphi$ and the direction of the interface. The other state developed at positive $k_y$ shows to have this asymmetry inverted.

\begin{figure*}[t]
 \includegraphics[width=0.9\textwidth]{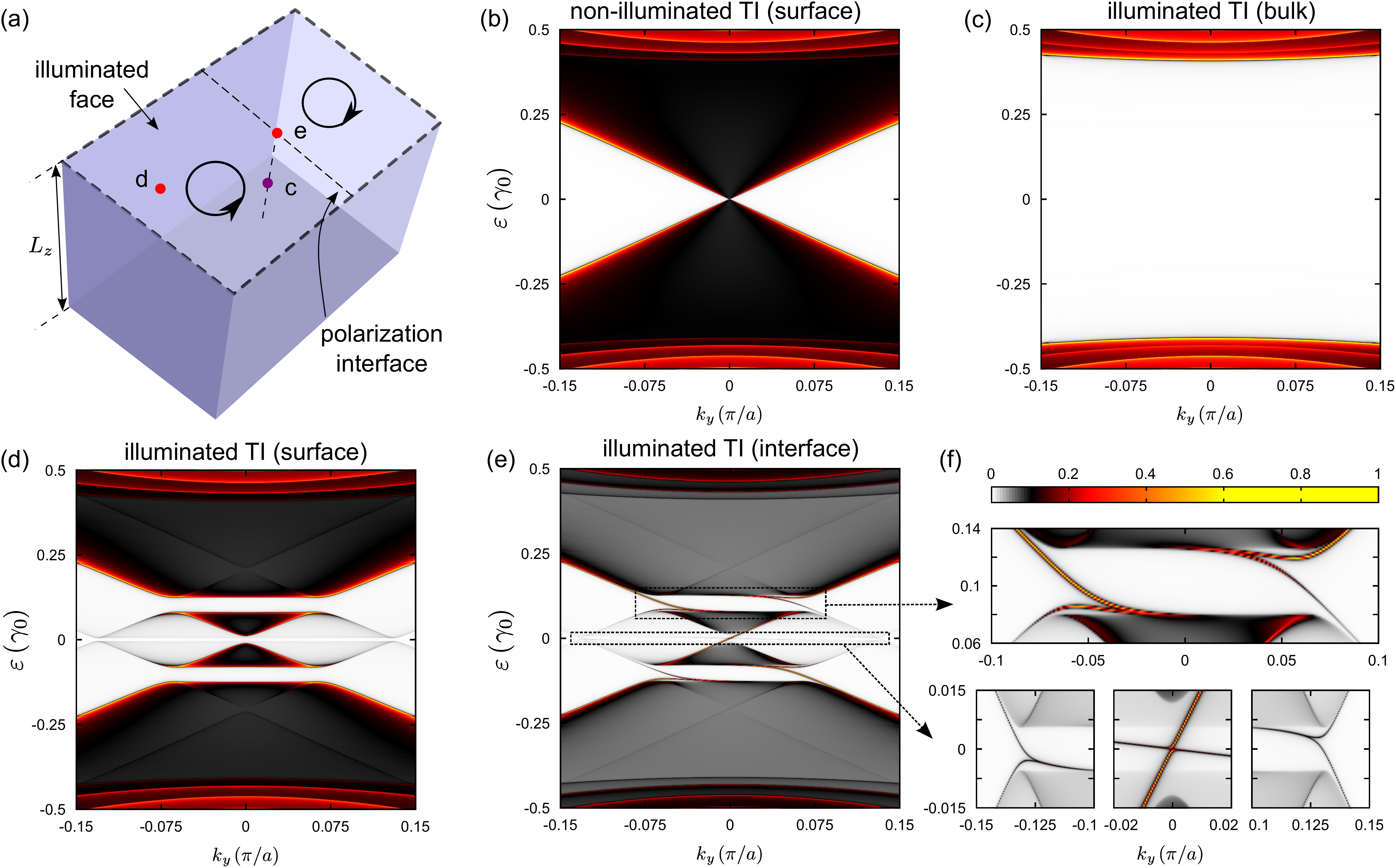}
 \caption{(color online). Normalized $\rho_{k_y}$ in an isotropic model for a 3D TI with $L_z = 9 a$: (a) Schematics of the considered setup indicating the different points at which $\rho_{k_y}$ is evaluated: (b) At the surface of a static TI. (c) Irradiated sample at the insulator's bulk ($z = 5 a$). (d) At the surface, away from the interface's region, here only one of the two lasers dominates. (e) At the surface, across the interface's region. (f) Used color scale in all plots and close-up of panel (e) at the ZB (top panel) and the ZC (bottom panels) regions. In all cases, the used parameters for the static Hamiltonian are $m_0 = 0.4$, $m_1 = 0.25$, and $m_2 = 0.5$~\cite{Chu2011}, and the laser parameters are 
$\hbar \Omega = 0.2 \gamma_0$, $\eta_0 = 2\pi a_0 a /\Phi_0 = 0.1$~\cite{note2}, $\varphi=\pi/2$, $x_0 = 3 a$, $\xi = 0.8$, $z_0 = a$ and includes the Floquet channels $m=\{-1,0,1\}$.}
 \label{fig:3}
\end{figure*} 

Using the solutions of Eq.~(\ref{eq:diffeq}) we calculate the time-averaged spin texture~\cite{Usaj2014} associated to the Floquet boundary states. Thanks to the spin-momentum locking present in the TI without radiation, there is a non-vanishing spin component in the Floquet states, $\braket{\braket{\sigma_x}} = 2 \sum_{m} \int \text{Re} [ \psi_{\uparrow,m}^{*}(x) \psi_{\downarrow,m}(x) ] \mathrm{d}x$, \textit{i.e.} the in-plane component perpendicular to the interface's direction. In all cases, $\braket{\braket{\sigma_x}}$ is proportional to the group velocity, as can be seen in Fig.~\ref{fig:2}. Since the direction of propagation can be tuned by the lasers, these robust states could be also interesting from the point of view of spin-polarized transport at a desired region of the TI's surface.


\textit{Three-dimensional lattice model and LDOS.--} Up to now our analysis is based on an effective 2D-model for the surface states of a TI. This poses the  question on whether these properties can be reproduced in a model accounting for the insulating bulk bands of a 3D TI. We therefore consider a lattice Hamiltonian which satisfies the four symmetries present in a strong TI~\cite{Zhang2012}. By taking a cubic lattice with parameter $a$, we obtain a tight-binding description for an isotropic TI~\cite{Chu2011}. The vector potential $\bm{A}(\bm{r},t) = \sum_{\tau=\pm} \bm{A}_\tau(z,t)f(\tau x)$, enters through Peierls' substitution as a time-dependent modulation of the hopping matrices coupling nearest neighbor sites. This accounts for (\textit{i}) a  gradual change $f(x)=[1+\exp(x/x_0)]^{-1}$ of the laser helicity, which produces a $\varphi$-oriented, linearly polarized field at the interface region, and (\textit{ii}) a photon absorption process across the layers of the TI which manifests through a decay in the laser intensity, $A_0(z) = a_0 \xi^{-z/z_0}$. In our simulations, $\xi$ and $z_0$ are adjusted in such a way that the laser becomes negligible at the bottom face of the irradiated sample. The resulting lattice Hamiltonian is derived in the Supplemental Material~\cite{note2}.

In Fig.~\ref{fig:3} we show the time-averaged $k_y$-resolved LDOS ($\rho_{k_y}$)~\cite{Piskunow2014,Usaj2014} in a geometry [see panel (a)] where the solid is infinite along the $x$ and $y$ directions, while in the $z$-direction it has $N_z = 9$ layers. A quantitative description is possible by adjusting the model parameters to, \textit{e.g.}, those estimated in Ref.~\cite{Zhang2009}, yielding larger values of $N_z$. In panel (b) we calculate $\rho_{k_y}$ for the non-illuminated material, where the gapless surface state crossing the bulk gap can be appreciated. Turning on the lasers, we evaluate $\rho_{k_y}$ at different points of the sample. Panel (c) shows  the sample's bulk region, where there is a clear absence of states within the insulating bulk gap. When moving to the top surface, to a region dominated by only one of the two lasers, panel (d) reveals the ZB and ZC gaps similar to those observed in Ref.~\cite{Wang2013}. Finally, once we arrive to the interface region, panel (e) shows the emergence of chiral states similar to those of Fig.~\ref{fig:2}. In the ZC region [bottom panels in Fig.~\ref{fig:3}(f)], we can observe that due to the small width of the gap the central state (forward mover) \textit{apparently} crosses it, reflecting the $-\tau/2$ contribution to $C_0$ from the renormalization of the $m=0$ states. Similar to Fig.~\ref{fig:2}(b) there is, however, a small admixture which hybridizes this forward mover state with the degenerate states around $k_y \sim 0$ (backward movers) and the final state crossing the gap shows a negative slope, as required by $\Delta C_0$. In this sense, the difference in the number of states with opposite direction of motion is again bounded to the calculated topological invariants $C_0$ and $C_1$ on each side of the interface and do not depend on its specific shape. The details of the wave functions and of the quasi-energy dispersion, however, do depend on the angle between the interface and the orientation $\varphi$ of the linearly polarized vector potential formed at that point~\cite{note2}.

\textit{Final remarks.--} In summary, we found that illuminating the surface of a 3D TI with a pair of counter-rotating lasers generate chiral, one-dimensional states confined at the interface region between the lasers. These states locate within the recently measured laser-induced gaps in ARPES~\cite{Wang2013}, for which we believe a small modification of the experimental setup would be enough for its observation. Additionally, these states have a finite time-averaged spin texture subjected to the spin-momentum locking effect of the bare material, making them interesting from the point of view of spin polarized transport. Our calculations in the low-energy regime are supported by simulations in a 3D lattice model, which accounts for the interface zone. Given the topological character of the Floquet bands, the qualitative properties of these interface states (chirality and spin-momentum locking) remain unaffected by the experimental details of the laser configuration (\textit{e.g.} fluctuations in their relative phase). Importantly, the existence of the topological states is not bounded to the local recovery of the time-reversal symmetry at the interface~\cite{note4}. Other choices in the setup including, \textit{e.g.}, the simultaneous irradiation of different faces of the TI, are of great interest and deserve further exploration since one could exploit different spin-textures and band curvatures~\cite{Zhang2012,Zhang2013} to achieve control over the chiral states.\\

\noindent
\textit{Acknowlegdments.--} We acknowledge financial support from SeCyT-UNC, ANPCyT (PICT
Bicentenario 2010-1060, PICT 2013-1045), CONICET (PIP 11220110100832) and SeCTyP-UNCuyo (06/C415), ICTP's associateship program (LEFFT and GU), the Alexander von Humboldt Foundation (LEFFT) and Simons Foundation (GU). \\   

\section{Supplemental Material}

In this supplemental material we provide additional details concerning: (a) The solutions of the differential equation (continuum model) presented in the main article, (b) the explicit form of the Floquet Hamiltonian for the lattice model of an isotropic 3D TI, (c) the role of the linear polarization angle $\varphi$ in the dispersion of the laser-induced interface states, and (d) the case of two lasers with different frequencies.\\

\begin{figure*}[tbh]
 \includegraphics[width=0.85\textwidth]{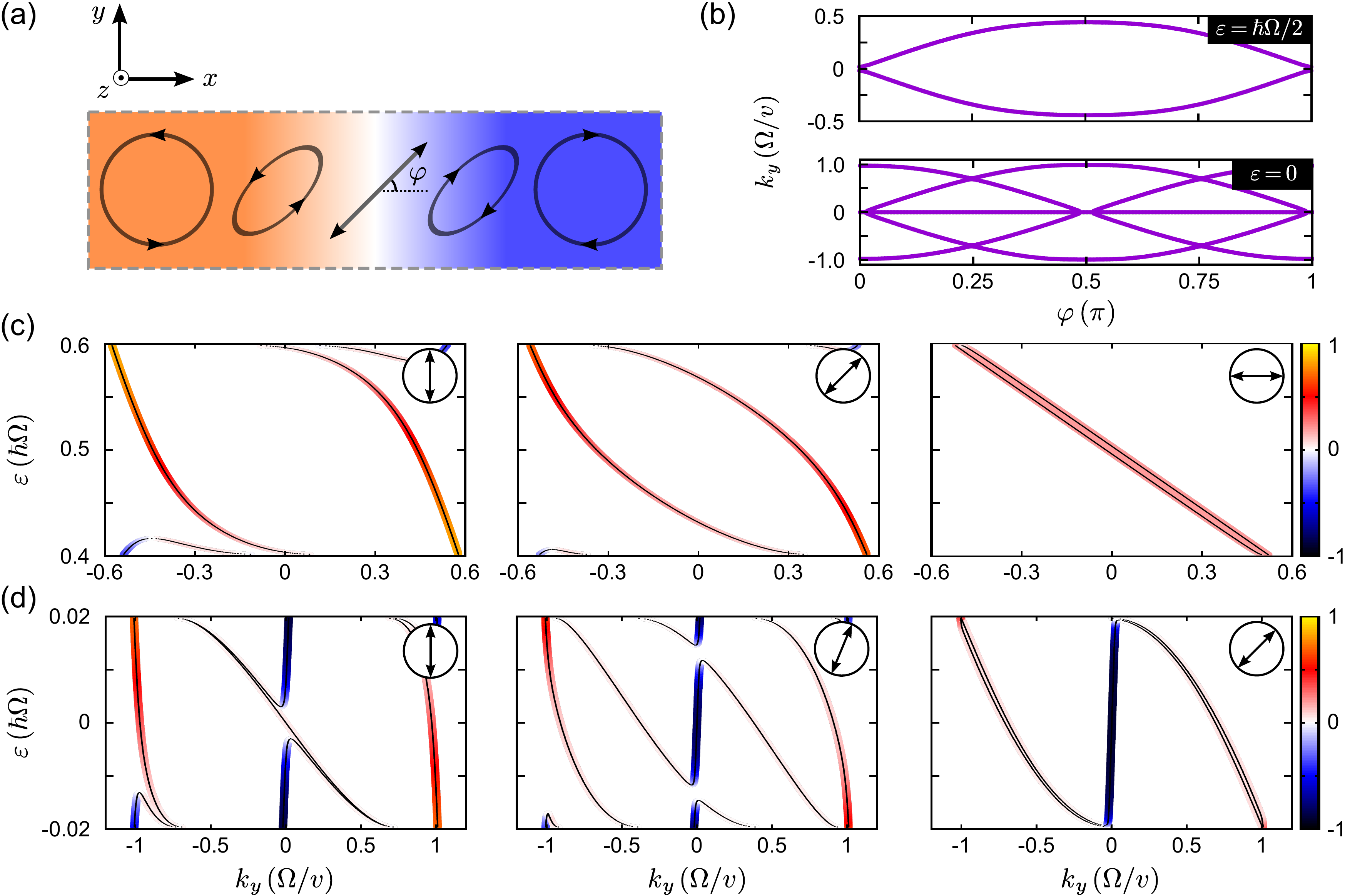}
 \caption{(color online). Interface states in the ZB and ZC gaps: (a) Schematics of the laser's rotation along the $x$-direction. In the vicinity of the interface region (white) the resulting linearly polarized field points in a direction $\varphi$ with respect to the $x$-axis. (b) Angular dependence of the interface states evaluated at the center of the ZB gap ($\varepsilon = \hbar\Omega/2$, top panel) and ZC gap ($\varepsilon = 0$, bottom panel). (c) ZB interface states evolution for $\varphi=\pi/2$ (left), $\varphi=\pi/4$ (center), and $\varphi=0$ (right). (d) ZC interface states evolution for $\varphi=\pi/2$ (left), $\varphi=3\pi/8$ (center), and $\varphi=\pi/4$ (right). In (c) and (d) the insets (arrows and circles) show the angle $\varphi$ and the color scales indicate the time-average $x$-component of spin $\braket{\braket{\sigma_x}}$ [cf. Eq.~(\ref{eq:spin})]. As in Fig.~\ref{fig:2}~ of the main text, we use $\gamma / \hbar\Omega = 0.2$.}
 \label{fig:S1}
\end{figure*}

\textbf{Solutions of the continuum model.--} The discussed differential equation in the main article for an interface region along the $y$-coordinate is of the form
\begin{equation}
\partial_x\boldsymbol{\Psi}(\bm{r}) = \mathcal{M}_{\tau(x)}\boldsymbol{\Psi}(\bm{r}),
\label{eq:diffeqS}
\end{equation}
where in general $\boldsymbol{\Psi}(\bm{r})=e^{i k_y}(...,\boldsymbol{\psi}_{-1},\boldsymbol{\psi}_{0},\boldsymbol{\psi}_{1},...)^\text{T}$, with $\boldsymbol{\psi}_m=(\psi_{\uparrow,m},\psi_{\downarrow,m})^\text{T}$, accounts for spin-up and spin-down states for each one of the $m$-Floquet channels and the coefficient matrix $\mathcal{M}_\tau = \text{i}\sigma_y (\mathcal{H}_F^\tau(k_y \bm{e}_y) -\varepsilon \mathcal{I})$, with $\mathcal{H}_F^\tau$ defined in Eq.~(\ref{eq:Floquet}), comes from the replacement $k_x \rightarrow -i \partial_x$ due to the broken traslational invariance. The solutions of Eq.~(\ref{eq:diffeqS}) can be obtained by diagonalizing $\mathcal{M}_\tau$ in the two regions ($x\gtrless0$) of the sample separately. These are of the form 
\begin{equation}
\psi_i(x) = \sum_j [\mathcal{U}_{\tau(x)}^{-1}]_{ij} C_j^{\tau(x)} e^{\lambda_j x},
\label{eq:solution}
\end{equation}
where $i$ labels the states of the truncated basis $\{\sigma,m\}$, $\mathcal{U}_\tau$ is the transformation matrix that diagonalizes $\mathcal{M}_\tau$ and $\tau(x) = -\sgn(x)$ determines the direction of rotation of the vector potentials at each side of the interface. For each one of the considered gapped regions, namely the zone center (ZC) and zone boundary (ZB), we work in a different truncation basis of Floquet channels to ensure a symmetric eigenvalue spectrum of $\mathcal{M}_\tau$ around zero. Specifically, we work in the Floquet space defined by the $m \in \{0,1\}$ for the ZB gap and $m \in \{-1,0,1\}$ for the ZC gap. This last guarantees that for an eigenvalue $\lambda_j$ of $\mathcal{M}_\tau$ there is always another $\lambda_k$ such that $\lambda_k = -\lambda_j$ and allows us to order them in the form $\text{Re}(\lambda_1)<...<\text{Re}(\lambda_{N/2})<0<\text{Re}(\lambda_{N/2+1})<...<\text{Re}(\lambda_N)$, with $N$ the dimension of the truncated space. Due to the specific form of the coefficient matrices $\mathcal{M}_\tau$, the eigenvalues $\lambda_j$ are independent of the $\tau$-index, yielding the same $\lambda$-spectrum in the two regions~\cite{note3}. To ensure the convergence of $\psi_i$ for $x\rightarrow \pm\infty$ in the considered gaps, we set to zero those coefficients $C_j^{\pm}$ which are associated to $\mathrm{Re}(\lambda_j) \lessgtr 0$. According to the above ordering, this implies
\begin{align}
C_j^{+} &= 0, &j &=1,...,N/2,\\
C_j^{-} &= 0, &j &=N/2+1,...,N.
\end{align}
The remaining coefficients are found by imposing a topological boundary condition~\cite{Zhang2012} across the interface between the two portions of the system. In the considered setup, this implies the continuity of $\boldsymbol{\Psi}(\bm{r})$ along the $x$-direction, where the sign in the Chern number is inverted, and it reads
\begin{equation}
\sum_{j=N/2+1}^N [\mathcal{U}_{+}^{-1}]_{ij} C_j^{+} = \sum_{j=1}^{N/2} [\mathcal{U}_{-}^{-1}]_{ij} C_j^{-}.
\end{equation}
Since in the above equation there is only one coefficient for each particular state $j$, we can define $\bm{C} = (C_1^{-},...,C_{N/2}^{-},C_{N/2+1}^{+},...,C_{N}^{+})^\mathrm{T}$, such that $\mathcal{Q}\bm{C} = \bm{0}$, with
\begin{equation}
[\mathcal{Q}]_{ij}=\left\{ 
\begin{array}{ll}
-[\mathcal{U}_{-}^{-1}]_{ij}, &j=1,...,N/2\\
+[\mathcal{U}_{+}^{-1}]_{ij}, &j=N/2+1,...,N.
\end{array}\right.
\label{eq:qmatrix}
\end{equation}
Through the condition $\det{\mathcal{Q}}=0$ we thus determine \textit{numerically} the energy $\varepsilon$ and momentum $k_y$ of the interface states within the ZC and ZB gaps where all the $\lambda$-eigenvalues have a non-vanishing real component (cf. Figs.~\ref{fig:2}~and \ref{fig:S1}).\\

\textit{Spin texture.--} Based on the above solutions for the interface state wave-functions we can now calculate the expectation values of the spin operator $\boldsymbol{\sigma}$. Let's assume we find a solution $\boldsymbol{\Psi}(\bm{r})$ such that it satisfies the characteristic equation $\det{\mathcal{Q}}=0$. Written in the Hilbert's real space ($\mathcal{R}$), the wave-function reads~\cite{Usaj2014}
\begin{equation}
\ket{\Psi(\bm{r},t)} = e^{i\varepsilon t} \sum_{\sigma,m} \psi_{\sigma,m}(x) e^{i k_y y} e^{i m \Omega t} 
\ket{\Psi_\sigma},
\end{equation}
where $\{\ket{\Psi_\sigma}\}$ is a complete basis for the spin states in $\mathcal{R}$ and the expectation value of the spin operator therefore reads
\begin{equation*}
\braket{\boldsymbol{\sigma}} = \sum_{\sigma,\sigma'} \sum_{m,m'} \int \mathrm{d}x \psi_{\sigma,m}^{*}(x) \psi_{\sigma',m'}(x) e^{-i(m-m')\Omega t}
[\boldsymbol{\sigma}]_{\sigma,\sigma'}.
\end{equation*}
For time-averaged quantities over a period $T = 2\pi/\Omega$ of the driving we observe that only the direct terms with $m' = m$ survive and hence
\begin{equation}
\braket{\braket{\boldsymbol{\sigma}}} = \sum_{\sigma,\sigma'} \sum_m \int \mathrm{d}x \psi_{\sigma,m}^{*}(x) \psi_{\sigma',m}(x) [\boldsymbol{\sigma}]_{\sigma,\sigma'}.
\label{eq:spin}
\end{equation}
This last was calculated in Figs.~\ref{fig:2}~and \ref{fig:S1} for the interface oriented along the $y$-direction and yields the same spin-momentum locking effect observed in static 3D TIs, \textit{i.e.}, $\braket{\braket{\sigma_y}} = \braket{\braket{\sigma_z}} = 0$ and finite $\braket{\braket{\sigma_x}}$.\\
  
\textbf{Lattice Floquet Hamiltonian.--} Here we derive the Floquet Hamiltonian for the lattice model introduced in the main article. We start from the static Hamiltonian for the isotropic TI of Ref.~\cite{Chu2011} for a cubic geometry with lattice parameter $a$:
\begin{equation}
H = \sum_{\bm{r}} c_{\bm{r}}^\dag M_0 c_{\bm{r}} + \sum_{\bm{r},\alpha} c_{\bm{r}}^\dag T_\alpha c_{\bm{r}+a\bm{e}_\alpha}+\text{H.c.},
\end{equation}
where the sum runs over the lattice's sites $\bm{r}$ and $\alpha=\{x,y,z\}$. The $4\times4$ on-site $M_0$ and hopping $T_\alpha$ matrices
\begin{equation}
M_0 = (m_0-6 m_1)\tau_z, \; T_\alpha = \left(m_1\tau_z-i\frac{m_2}{2} \tau_x \sigma_\alpha \right),
\end{equation}
account for both the orbital and spin degrees of freedom through the Pauli's matrices $\tau_\alpha$ and $\sigma_\alpha$, respectively. Here $m_0$, $m_1$ and $m_2$ are standard parameters of the model~\cite{Chu2011} whose scale is set by the hopping term $\gamma_0$. The driving field is included by Peierls' substitution as a phase-modulation of the hopping amplitudes coupling nearest-neigbour sites. This incorporates space and time dependencies in $T_\alpha$ through the vector potential, defined as $\bm{A}(\bm{r},t) = \bm{A}_{-}(z,t)f(x)+\bm{A}_{+}(z,t)(1-f(x))$, where $f(x)=1/[1+\exp(x/x_0)]$ and
\begin{equation}
\bm{A}_\tau = A_0(z)\left[ \cos(\tau\Omega t+\varphi) \, \bm{e}_x + \sin(\tau\Omega t+\varphi)\, \bm{e}_y \right],
\end{equation}
with $A_0(z)=a_0 \xi^{z/z_0}$ the magnitude of the vector potential, assumed to be attenuated along $z$ due to absorption within the solid. At the interface region ($x=0$) the two lasers add-up, yielding a linearly polarized field $\bm{A}= A_0(z) \cos(\Omega t) \bm{e}_\varphi$ whose direction forms an angle $\varphi$ with respect to the $x$-axis. The phase-modulation thus enters through the line integrals
\begin{equation}
T_\alpha(\bm{r},t) \rightarrow T_\alpha \exp \left[i \frac{2\pi}{\Phi_0} \int_{\bm{r}}^{\bm{r}+a\bm{e}_\alpha} \bm{A}(\bm{r},t) \cdot \text{d}\bm{\ell}_\alpha \right],
\end{equation}
with $\Phi_0$ the magnetic flux quantum. Now that we have the explicit time-dependence of the hopping matrices, the Floquet Hamiltonian can be easily obtained by taking the Fourier components of the above phase-modulation, i.e., $[\mathcal{H}_F]_{n,m}=\frac{1}{T}\int_0^T H(t)e^{i(n-m)\Omega t} \text{d}t$, such that
\begin{equation}
\mathcal{H}_F = \sum_{\bm{r}} c_{\bm{r}}^\dag \mathcal{M}_0 c_{\bm{r}} + \sum_{\bm{r},\alpha} c_{\bm{r}}^\dag \mathcal{T}_\alpha(\bm{r}) c_{\bm{r}+a\bm{e}_\alpha}+\text{H.c.}
\label{eq:red}
\end{equation}
The above static matrices $M_0$ and $T_\alpha$ generalize in Floquet space to
\begin{equation}
\mathcal{M}_0 = M_0 \otimes I + I \otimes N_\Omega, \; \mathcal{T}_\alpha(\bm{r}) = T_\alpha \otimes \gamma_{\alpha}(\bm{r}),
\end{equation}
where $[N_\Omega]_{n,m} = n\hbar\Omega \delta_{n,m}$. The tensor product follows from the $\mathcal{R} \otimes \mathcal{T}$ structure of the Floquet space and $[\gamma_\alpha(\bm{r})]_{n,m} = \gamma_\alpha^{n-m}(\bm{r})$, with 
\begin{eqnarray}
\gamma_x^n &=& \sum_m i^n e^{i(2m-n)\varphi} J_m(\eta g) J_{n-m}(\eta (1-g)),\\
\gamma_y^n &=& \sum_m e^{i(2m-n)\varphi} J_m(\eta f) J_{n-m}(\eta (f-1)),
\end{eqnarray}
and $\gamma_z^n=\delta_{n,0}$. Here, $J_n$ is the $n$-th order Bessel function of the first kind, $g(x) = \frac{1}{a}\int_x^{x+a} f(x') \text{d}x'$ and $\eta(z) = 2 \pi a A_0(z)/\Phi_0$, such that $\eta_0 = 2\pi a_0 a /\Phi_0$ sets the strength of the couplings between the Floquet channels.\\

\textbf{Polarization angle.--} In this section we describe in detail the role of the angle $\varphi$ of the linearly polarized laser formed in the vicinity of the interface region [see Figs.~\ref{fig:1}~and~\ref{fig:S1}(a)].

We start our description based on the solutions discussed above for the low-energy, continuum model. In this case, we assumed a sudden change in the helicity of the laser through $\tau(x) = -\sgn(x)$ [see Eq.~(\ref{eq:solution})], and the relevant angular dependence comes from the sum between the two contributions, \textit{i.e.}, $\bm{A}_+(t)+\bm{A}_-(t) = A_0 \cos (\Omega t) \bm{e}_\varphi$, which yields a linearly polarized field whose oscillation direction forms an angle $\varphi$ with respect to the $x$-axis. Although this linearly polarized field is not explicitly present in the continuum model, the dispersions of the interface states strongly depend on the relative orientation of the two fields which is kept fixed by $\varphi$ at any value of time.

In Fig.~\ref{fig:S1} we show the solutions to $\det \mathcal{Q} = 0$ discussed above in the vicinity of the ZC and ZB gaps. In panel (b) we fix the energy at the center of each one of the gaps and evaluate the $k_y$-position of the interface states for arbitrary orientation $\varphi$. It can be seen that for the ZB region (top panel) these states are $\pi$-periodic while in the ZC region (bottom panel) they turn to be $\pi/2$-periodic. As we will discuss below, the difference in the $\varphi$-periodicity of the two regions bears a direct resemblance with the Floquet spectrum in a sample irradiated with a linearly polarized field. In panels (c) and (d) we explore the dispersion relations of the interface states for intermediate angles, where the dependence with $\varphi$ becomes evident. For the ZB gap, cf. Fig.~\ref{fig:S1}(c), the two interface states move backwards with respect to the interface's direction $y$. These start at maximum separation for $\varphi=\pi/2$ (left panel) and then they shift to the center $k_y \sim 0$, becoming almost degenerate for $\varphi=0$ (right panel). In this $\varphi$-evolution we observe, in addition, a change in the group velocities, and together with it a decreasing spin polarization. For the ZC gap [see panel (d)], we notice a central state moving forwards and dominated by a negative spin component. This state remains almost in the same $k_y$-position regardless the value of $\varphi$. The other states (backward movers) shift in such a way that those in the center are degenerate for $\varphi=\pi/2$ (left panel) and then they merge with those initially placed in the extremes for $\varphi=\pi/4$ (right panel). This marked difference between forward and backward movers has indeed its origin in the different contributions to the Chern number $C_0 = -\tau/2+2\tau$ discussed in the main article. These contributions are not, however, completely `separated' when including the interface boundary, manifesting itself as an avoided crossing between forward and backward movers. In any case, this admixture between states with different directions of propagation is in agreement with the bulk-boundary correspondence, since this last dictates not the total number of interface states but the difference between forward and backward movers.

\begin{figure*}[tbh]
 \includegraphics[width=0.95\textwidth]{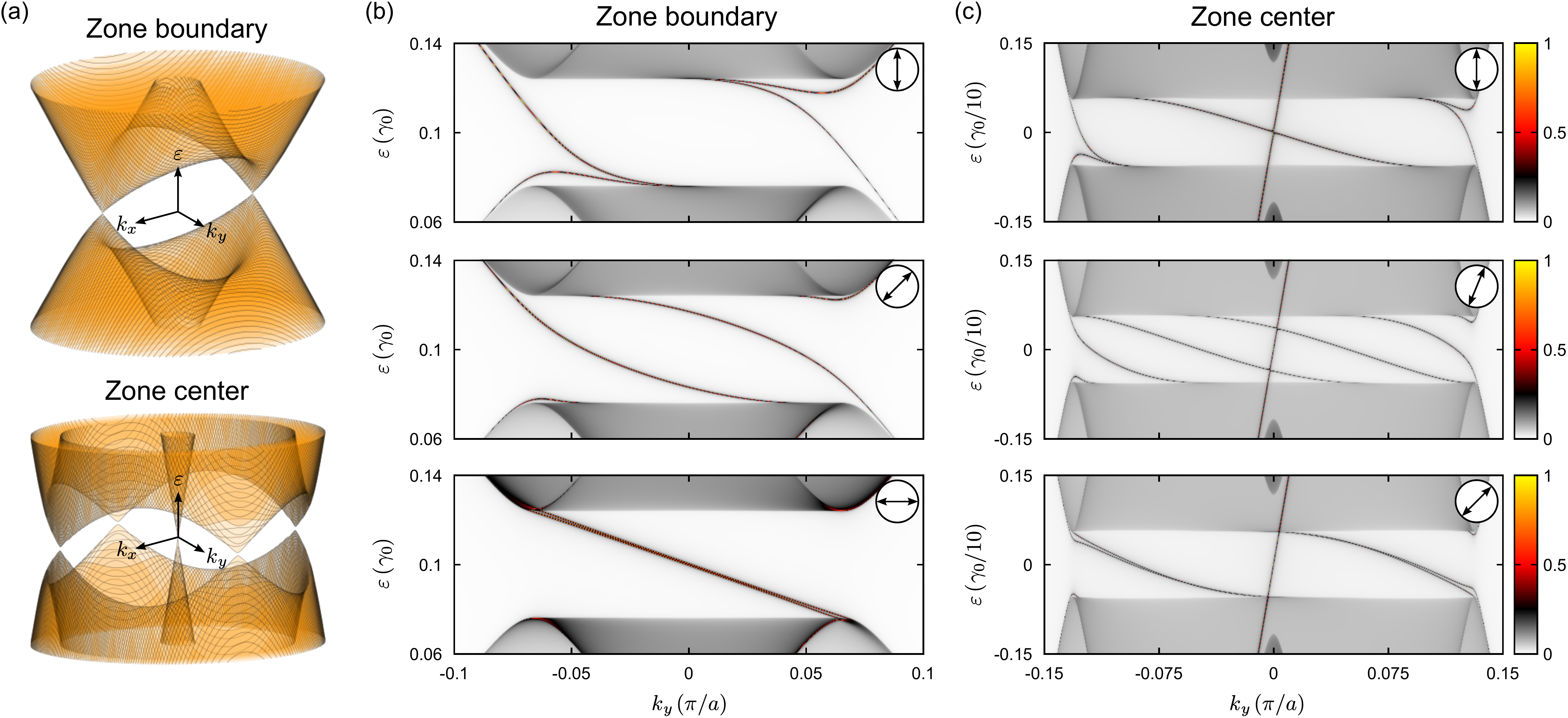}
 \caption{(color online). (a) Typical dispersion relations for the irradiated system with linearly polarized field around the ZB (top) and ZC (bottom) regions. (b)--(c) Normalized $\rho_{k_y}$ for the isotropic lattice model in Eq.~(\ref{eq:red}): (b) In the vicinity of the ZB gap, for $\varphi=\pi/2$ (top), $\varphi=\pi/4$ (center), and $\varphi=0$ (bottom). (b) $\rho_{k_y}$ around the ZC gap for $\varphi=\pi/2$ (top), $\varphi=3\pi/8$ (center), and $\varphi=\pi/4$ (bottom). The remaining parameters coincide with those used in Fig.~\ref{fig:3}~of the main article: $N_z=9$, $m_0 = 0.4$, $m_1 = 0.25$, $m_2 = 0.5$, $\hbar \Omega = 0.2 \gamma_0$, $\eta_0 = 2\pi a_0 a /\Phi_0 = 0.1$, $x_0 = 3 a$, $\xi = 0.8$ and $z_0 = a$.}
 \label{fig:S2}
\end{figure*}

To elucidate the $k_y$-position of the interface states in the above discussed dispersion relations, we show in Fig.~\ref{fig:S2}(a) the calculated Floquet spectrum of a sample which is being irradiated by a \text{single} laser with linear polarization~\cite{Fregoso2013}. In this setup both the traslational invariance along $x$ and the time-reversal symmetry are recovered. The corresponding Floquet Hamiltonian in this case writes:
\begin{equation}
\mathcal{H}_F = H_0\otimes I + I \otimes N_\Omega + \frac{\gamma}{2}(\sin\varphi \sigma_x - \cos\varphi \sigma_y) \otimes \Delta, 
\label{eq:lineal}
\end{equation}
where $H_0 = \hbar v (k_y \sigma_x - k_x \sigma_y)$, $\gamma = v e A_0/c$ and $[\Delta]_{m,n}=\delta_{m,n+1}+\delta_{m,n-1}$. From Fig.~\ref{fig:S2}(a) it can be seen that the aforementioned laser-induced gaps now close at certain points as a consequence of the reinstated time-reversal symmetry. This `closing' of the gaps occurs at two points in the ZB region (top) while in the ZC region (bottom) we have a central Dirac cone (gapless) surrounded by four closing points in a different shell of the spectrum. From Eq.~(\ref{eq:lineal}) we notice that the role of the polarization angle $\varphi$ is to rotate the whole spectrum and with it the points where the gaps close. Returning to the initial setup with the two counter-rotating, circularly polarized lasers, we can now associate the $k_y$-position of the interface states bridging the gaps to these points in which the gaps close when accounting for the linear polarization scenario. In particular, in the ZC region the difference between forward and backward movers can be attributed to the shells of the spectrum to which they belong: The forward mover is tied-up to the central Dirac cone, related to the $m=0$ channel, while the backward movers are placed in the outer closing points, associated to the crossing between $m=-1$ and $m=1$ cones. Within this correspondence between interface states and the points where the gaps close, it is possible to understand the above periodicities in both the ZB and ZC gaps [see Fig.~\ref{fig:S1}(b)] as a consequence of the rigid rotation of the spectrum when sweeping $\varphi$.

To further support the above discussed $\varphi$-dependence in the dispersion relations of the interface states, we also present calculations of the time-averaged LDOS ($\rho_{k_y}$) evaluated along the interface region for the lattice model described in the previous section. In Figs.~\ref{fig:S2}(b) and (c) we show the normalized $\rho_{k_y}$ in the ZB and ZC gap regions for the same angles $\varphi$ as in Figs.~\ref{fig:S1}(c) and (d), respectively. All the discussed effects are reproduced qualitatively. In this case where the intensity of the laser is smaller than the one used in the continuum model, the mixing between forward and backward movers becomes almost unnoticeable.\\

\textbf{Lasers with different frequencies.--} As a concrete test example to inspect the robustness of the laser-induced interface states against possible breaking of the local time-reversal symmetry (TRS) we consider the case of two laser beams defined with two different frequencies and opposite helicities. To ensure the global periodicity of the radiation field, we take commensurable frequencies. Indeed, the TRS is broken everywhere in this system.

\begin{figure*}[tbh]
 \includegraphics[width=0.95\textwidth]{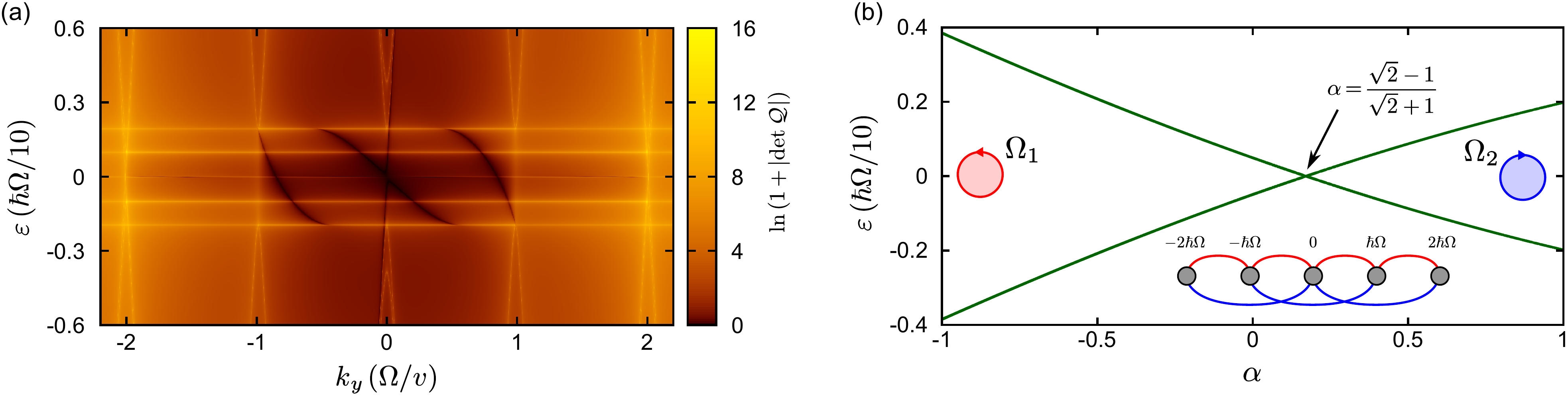}
 \caption{(color online). (a) Solution map of Eq.~(\ref{eq:diffeqS}) [\textit{i.e.}, $\ln(1+|\det \mathcal{Q}|)$] for the case of two laser beams with different frequencies. The used parameters are: $\Omega_1 = \Omega$, $\Omega_2 = 2\Omega$, $\gamma/\hbar \Omega = 0.2$ and $\varphi=0$. (b) Zone center gap closing as a function of the relative strength $\alpha$ between the lasers. In the limit case $\alpha = -1$ ($+1$) the system is being irradiated with a single laser of frequency $\Omega_1$ ($\Omega_2$) and helicity $\tau = +1$ ($-1$). The parameters coincide with those in panel (a). Inset: Scheme of the considered Floquet Hamiltonian indicating the different replicas and their couplings given by the lasers. The lines involve a photon absorption/emission of energy $\hbar \Omega$ (red lines) or $2\hbar \Omega$ (blue lines) associated to each one of the lasers.}
 \label{fig:S3}
\end{figure*}

In Fig.~\ref{fig:S3}(a) we calculate the determinant of the $\mathcal{Q}$-matrix as defined in Eq.~(\ref{eq:qmatrix}) around the Floquet zone center (ZC) region. The solutions to Eq.~(\ref{eq:diffeqS}) correspond to those points (black traces) where the map goes to zero. Here, we can appreciate the presence and chirality of the laser-induced interface states for the case of two laser beams with frequencies $\Omega_1=\Omega$ and $\Omega_2 = 2\Omega$, respectively, and opposite helicities. We have tested other cases with smaller frequency ratios, and they all show the same features but become technically demanding~\cite{note5}. The spectrum results similar to that of the single frequency case (cf. Fig.~\ref{fig:2}(b) in the main manuscript). At very low energies, differences are expected from this situation regarding the effect of additional Fourier replicas that need to be included in our calculation. These, however, are well-understood and follow the hierarchy recently introduced in Ref.~[\onlinecite{Piskunow2015}].

We emphasize that the formation of chiral interface states even in this system with fully broken TRS can be explained by the closing of the laser-induced ZC gap when connecting two Hamiltonians of different topology (winding number) in a continuous manner. To verify this, we compute the Floquet-Bloch bands of the Hamiltonian 
\begin{equation}
H(t) = \frac{1-\alpha}{2} H_{+}(\Omega_1,t)+\frac{1+\alpha}{2} H_{-}(\Omega_2,t),
\end{equation}
where $H_\tau(\Omega,t)$ is the Hamiltonian describing the surface of the topological insulator in the presence of a circularly polarized laser of frequency $\Omega$ whose helicity is determined by $\tau=\pm 1$. Here, the constant $\alpha$ simply parametrizes the continuous change of $H(t)$ from $H_{+}(\Omega_1,t)$ to $H_{-}(\Omega_2,t)$. In the case described in the main manuscript, this parameter would depend on the coordinate perpendicular to the interface and represents the relative strength between the two laser beams. Here, however, it is an interpolation parameter and thus $H(t)$ can be regarded as a local Hamiltonian describing the interface region. By taking a particular symmetry point ($\mathbf{k}=0$), we find that when $\alpha$ goes from -1 to 1, hence connecting the two topological phases, the ZC gap necessarily closes as indicated in Fig.~\ref{fig:S3}(b). This occurs for the value of $\alpha$ indicated by the arrow when the replicas from $m=-2$ to $m=2$ are retained (the specific value of $\alpha$ where the gap closes shows little change when considering more replicas). Although not shown, the ZC gap also closes in other $\mathbf{k}$-points as $\alpha$ is swept. Thus, the gap closing at certain points in the $k$-space as connecting the two topological distinct phases ensures bulk-boundary correspondence to hold in this system. Very importantly, TRS is always broken regardless the value of $\alpha$.

\bibliographystyle{apsrev4-1_title}
\bibliography{cite,notes}

\begin{thebibliography}{63}%
\makeatletter
\providecommand \@ifxundefined [1]{%
 \@ifx{#1\undefined}
}%
\providecommand \@ifnum [1]{%
 \ifnum #1\expandafter \@firstoftwo
 \else \expandafter \@secondoftwo
 \fi
}%
\providecommand \@ifx [1]{%
 \ifx #1\expandafter \@firstoftwo
 \else \expandafter \@secondoftwo
 \fi
}%
\providecommand \natexlab [1]{#1}%
\providecommand \enquote  [1]{``#1''}%
\providecommand \bibnamefont  [1]{#1}%
\providecommand \bibfnamefont [1]{#1}%
\providecommand \citenamefont [1]{#1}%
\providecommand \href@noop [0]{\@secondoftwo}%
\providecommand \href [0]{\begingroup \@sanitize@url \@href}%
\providecommand \@href[1]{\@@startlink{#1}\@@href}%
\providecommand \@@href[1]{\endgroup#1\@@endlink}%
\providecommand \@sanitize@url [0]{\catcode `\\12\catcode `\$12\catcode
  `\&12\catcode `\#12\catcode `\^12\catcode `\_12\catcode `\%12\relax}%
\providecommand \@@startlink[1]{}%
\providecommand \@@endlink[0]{}%
\providecommand \url  [0]{\begingroup\@sanitize@url \@url }%
\providecommand \@url [1]{\endgroup\@href {#1}{\urlprefix }}%
\providecommand \urlprefix  [0]{URL }%
\providecommand \Eprint [0]{\href }%
\providecommand \doibase [0]{http://dx.doi.org/}%
\providecommand \selectlanguage [0]{\@gobble}%
\providecommand \bibinfo  [0]{\@secondoftwo}%
\providecommand \bibfield  [0]{\@secondoftwo}%
\providecommand \translation [1]{[#1]}%
\providecommand \BibitemOpen [0]{}%
\providecommand \bibitemStop [0]{}%
\providecommand \bibitemNoStop [0]{.\EOS\space}%
\providecommand \EOS [0]{\spacefactor3000\relax}%
\providecommand \BibitemShut  [1]{\csname bibitem#1\endcsname}%
\let\auto@bib@innerbib\@empty
\bibitem [{\citenamefont {Novoselov}\ \emph {et~al.}(2005)\citenamefont
  {Novoselov}, \citenamefont {Geim}, \citenamefont {Morozov}, \citenamefont
  {Jiang}, \citenamefont {Katsnelson}, \citenamefont {Grigorieva},
  \citenamefont {Dubonos},\ and\ \citenamefont {Firsov}}]{Novoselov2005a}%
  \BibitemOpen
  \bibfield  {author} {\bibinfo {author} {\bibfnamefont {K.~S.}\ \bibnamefont
  {Novoselov}}, \bibinfo {author} {\bibfnamefont {A.~K.}\ \bibnamefont {Geim}},
  \bibinfo {author} {\bibfnamefont {S.~V.}\ \bibnamefont {Morozov}}, \bibinfo
  {author} {\bibfnamefont {D.}~\bibnamefont {Jiang}}, \bibinfo {author}
  {\bibfnamefont {M.~I.}\ \bibnamefont {Katsnelson}}, \bibinfo {author}
  {\bibfnamefont {I.~V.}\ \bibnamefont {Grigorieva}}, \bibinfo {author}
  {\bibfnamefont {S.~V.}\ \bibnamefont {Dubonos}}, \ and\ \bibinfo {author}
  {\bibfnamefont {A.~A.}\ \bibnamefont {Firsov}},\ }\bibfield  {title}
  {\enquote {\bibinfo {title} {Two-dimensional gas of massless dirac fermions
  in graphene},}\ }\href {http://dx.doi.org/10.1038/nature04233} {\bibfield
  {journal} {\bibinfo  {journal} {Nature}\ }\textbf {\bibinfo {volume} {438}},\
  \bibinfo {pages} {197} (\bibinfo {year} {2005})}\BibitemShut {NoStop}%
\bibitem [{\citenamefont {Zhang}\ \emph {et~al.}(2005)\citenamefont {Zhang},
  \citenamefont {Tan}, \citenamefont {Stormer},\ and\ \citenamefont
  {Kim}}]{Zhang2005}%
  \BibitemOpen
  \bibfield  {author} {\bibinfo {author} {\bibfnamefont {Y.}~\bibnamefont
  {Zhang}}, \bibinfo {author} {\bibfnamefont {Y.-W.}\ \bibnamefont {Tan}},
  \bibinfo {author} {\bibfnamefont {H.~L.}\ \bibnamefont {Stormer}}, \ and\
  \bibinfo {author} {\bibfnamefont {P.}~\bibnamefont {Kim}},\ }\bibfield
  {title} {\enquote {\bibinfo {title} {Experimental observation of the quantum
  hall effect and berry's phase in graphene},}\ }\href
  {http://dx.doi.org/10.1038/nature04235} {\bibfield  {journal} {\bibinfo
  {journal} {Nature}\ }\textbf {\bibinfo {volume} {438}},\ \bibinfo {pages}
  {201} (\bibinfo {year} {2005})}\BibitemShut {NoStop}%
\bibitem [{\citenamefont {Geim}\ and\ \citenamefont
  {Novoselov}(2007)}]{Geim2007}%
  \BibitemOpen
  \bibfield  {author} {\bibinfo {author} {\bibfnamefont {A.~K.}\ \bibnamefont
  {Geim}}\ and\ \bibinfo {author} {\bibfnamefont {K.~S.}\ \bibnamefont
  {Novoselov}},\ }\bibfield  {title} {\enquote {\bibinfo {title} {The rise of
  graphene},}\ }\href {http://dx.doi.org/10.1038/nmat1849} {\bibfield
  {journal} {\bibinfo  {journal} {Nature Materials}\ }\textbf {\bibinfo
  {volume} {6}},\ \bibinfo {pages} {183} (\bibinfo {year} {2007})}\BibitemShut
  {NoStop}%
\bibitem [{\citenamefont {K\"onig}\ \emph {et~al.}(2007)\citenamefont
  {K\"onig}, \citenamefont {Wiedmann}, \citenamefont {Br\"une}, \citenamefont
  {Roth}, \citenamefont {Buhmann}, \citenamefont {Molenkamp}, \citenamefont
  {Qi},\ and\ \citenamefont {Zhang}}]{Koenig2007}%
  \BibitemOpen
  \bibfield  {author} {\bibinfo {author} {\bibfnamefont {M.}~\bibnamefont
  {K\"onig}}, \bibinfo {author} {\bibfnamefont {S.}~\bibnamefont {Wiedmann}},
  \bibinfo {author} {\bibfnamefont {C.}~\bibnamefont {Br\"une}}, \bibinfo
  {author} {\bibfnamefont {A.}~\bibnamefont {Roth}}, \bibinfo {author}
  {\bibfnamefont {H.}~\bibnamefont {Buhmann}}, \bibinfo {author} {\bibfnamefont
  {L.~W.}\ \bibnamefont {Molenkamp}}, \bibinfo {author} {\bibfnamefont {X.-L.}\
  \bibnamefont {Qi}}, \ and\ \bibinfo {author} {\bibfnamefont {S.-C.}\
  \bibnamefont {Zhang}},\ }\bibfield  {title} {\enquote {\bibinfo {title}
  {Quantum spin hall insulator state in hgte quantum wells},}\ }\href
  {http://www.sciencemag.org/content/318/5851/766.abstrac} {\bibfield
  {journal} {\bibinfo  {journal} {Science}\ }\textbf {\bibinfo {volume}
  {318}},\ \bibinfo {pages} {766} (\bibinfo {year} {2007})}\BibitemShut
  {NoStop}%
\bibitem [{\citenamefont {Hsieh}\ \emph {et~al.}(2008)\citenamefont {Hsieh},
  \citenamefont {Qian}, \citenamefont {Wray}, \citenamefont {Xia},
  \citenamefont {Hor}, \citenamefont {Cava},\ and\ \citenamefont
  {Hasan}}]{Hsieh2008}%
  \BibitemOpen
  \bibfield  {author} {\bibinfo {author} {\bibfnamefont {D.}~\bibnamefont
  {Hsieh}}, \bibinfo {author} {\bibfnamefont {D.}~\bibnamefont {Qian}},
  \bibinfo {author} {\bibfnamefont {L.}~\bibnamefont {Wray}}, \bibinfo {author}
  {\bibfnamefont {Y.}~\bibnamefont {Xia}}, \bibinfo {author} {\bibfnamefont
  {Y.~S.}\ \bibnamefont {Hor}}, \bibinfo {author} {\bibfnamefont {R.~J.}\
  \bibnamefont {Cava}}, \ and\ \bibinfo {author} {\bibfnamefont {M.~Z.}\
  \bibnamefont {Hasan}},\ }\bibfield  {title} {\enquote {\bibinfo {title} {A
  topological dirac insulator in a quantum spin hall phase},}\ }\href
  {http://dx.doi.org/10.1038/nature06843} {\bibfield  {journal} {\bibinfo
  {journal} {Nature}\ }\textbf {\bibinfo {volume} {452}},\ \bibinfo {pages}
  {970} (\bibinfo {year} {2008})}\BibitemShut {NoStop}%
\bibitem [{\citenamefont {Kane}\ and\ \citenamefont {Mele}(2005)}]{Kane2005}%
  \BibitemOpen
  \bibfield  {author} {\bibinfo {author} {\bibfnamefont {C.~L.}\ \bibnamefont
  {Kane}}\ and\ \bibinfo {author} {\bibfnamefont {E.~J.}\ \bibnamefont
  {Mele}},\ }\bibfield  {title} {\enquote {\bibinfo {title} {Quantum spin hall
  effect in graphene},}\ }\href
  {http://link.aps.org/doi/10.1103/PhysRevLett.95.226801} {\bibfield  {journal}
  {\bibinfo  {journal} {Phys. Rev. Lett.}\ }\textbf {\bibinfo {volume} {95}},\
  \bibinfo {pages} {226801} (\bibinfo {year} {2005})}\BibitemShut {NoStop}%
\bibitem [{\citenamefont {Bernevig}\ \emph {et~al.}(2006)\citenamefont
  {Bernevig}, \citenamefont {Hughes},\ and\ \citenamefont
  {Zhang}}]{Bernevig2006}%
  \BibitemOpen
  \bibfield  {author} {\bibinfo {author} {\bibfnamefont {B.~A.}\ \bibnamefont
  {Bernevig}}, \bibinfo {author} {\bibfnamefont {T.~L.}\ \bibnamefont
  {Hughes}}, \ and\ \bibinfo {author} {\bibfnamefont {S.-C.}\ \bibnamefont
  {Zhang}},\ }\bibfield  {title} {\enquote {\bibinfo {title} {Quantum spin hall
  effect and topological phase transition in hgte quantum wells},}\ }\href
  {http://www.sciencemag.org/content/314/5806/1757.abstract N2 - We show that
  the quantum spin Hall (QSH) effect, a state of matter with topological
  properties distinct from those of conventional insulators, can be realized in
  mercury telluride-cadmium telluride semiconductor quantum wells. When the
  thickness of the quantum well is varied, the electronic state changes from a
  normal to an “inverted” type at a critical thickness dc. We show that
  this transition is a topological quantum phase transition between a
  conventional insulating phase and a phase exhibiting the QSH effect with a
  single pair of helical edge states. We also discuss methods for experimental
  detection of the QSH effect.} {\bibfield  {journal} {\bibinfo  {journal}
  {Science}\ }\textbf {\bibinfo {volume} {314}},\ \bibinfo {pages} {1757}
  (\bibinfo {year} {2006})}\BibitemShut {NoStop}%
\bibitem [{\citenamefont {Fu}\ and\ \citenamefont {Kane}(2007)}]{Fu2007}%
  \BibitemOpen
  \bibfield  {author} {\bibinfo {author} {\bibfnamefont {L.}~\bibnamefont
  {Fu}}\ and\ \bibinfo {author} {\bibfnamefont {C.~L.}\ \bibnamefont {Kane}},\
  }\bibfield  {title} {\enquote {\bibinfo {title} {Topological insulators with
  inversion symmetry},}\ }\href {\doibase 10.1103/PhysRevB.76.045302}
  {\bibfield  {journal} {\bibinfo  {journal} {Phys. Rev. B}\ }\textbf {\bibinfo
  {volume} {76}},\ \bibinfo {pages} {045302} (\bibinfo {year}
  {2007})}\BibitemShut {NoStop}%
\bibitem [{\citenamefont {Hasan}\ and\ \citenamefont {Kane}(2010)}]{Hasan2010}%
  \BibitemOpen
  \bibfield  {author} {\bibinfo {author} {\bibfnamefont {M.~Z.}\ \bibnamefont
  {Hasan}}\ and\ \bibinfo {author} {\bibfnamefont {C.~L.}\ \bibnamefont
  {Kane}},\ }\bibfield  {title} {\enquote {\bibinfo {title}
  {\textit{Colloquium} : Topological insulators},}\ }\href {\doibase
  10.1103/RevModPhys.82.3045} {\bibfield  {journal} {\bibinfo  {journal} {Rev.
  Mod. Phys.}\ }\textbf {\bibinfo {volume} {82}},\ \bibinfo {pages} {3045}
  (\bibinfo {year} {2010})}\BibitemShut {NoStop}%
\bibitem [{\citenamefont {Moore}(2010)}]{Moore2010}%
  \BibitemOpen
  \bibfield  {author} {\bibinfo {author} {\bibfnamefont {J.~E.}\ \bibnamefont
  {Moore}},\ }\bibfield  {title} {\enquote {\bibinfo {title} {The birth of
  topological insulators},}\ }\href {http://dx.doi.org/10.1038/nature08916}
  {\bibfield  {journal} {\bibinfo  {journal} {Nature}\ }\textbf {\bibinfo
  {volume} {464}},\ \bibinfo {pages} {194} (\bibinfo {year}
  {2010})}\BibitemShut {NoStop}%
\bibitem [{\citenamefont {Bonaccorso}\ \emph {et~al.}(2010)\citenamefont
  {Bonaccorso}, \citenamefont {Sun}, \citenamefont {Hasan},\ and\ \citenamefont
  {Ferrari}}]{Bonaccorso2010}%
  \BibitemOpen
  \bibfield  {author} {\bibinfo {author} {\bibfnamefont {F.}~\bibnamefont
  {Bonaccorso}}, \bibinfo {author} {\bibfnamefont {Z.}~\bibnamefont {Sun}},
  \bibinfo {author} {\bibfnamefont {T.}~\bibnamefont {Hasan}}, \ and\ \bibinfo
  {author} {\bibfnamefont {A.~C.}\ \bibnamefont {Ferrari}},\ }\bibfield
  {title} {\enquote {\bibinfo {title} {Graphene photonics and
  optoelectronics},}\ }\href {http://dx.doi.org/10.1038/nphoton.2010.186}
  {\bibfield  {journal} {\bibinfo  {journal} {Nat Photon}\ }\textbf {\bibinfo
  {volume} {4}},\ \bibinfo {pages} {611} (\bibinfo {year} {2010})}\BibitemShut
  {NoStop}%
\bibitem [{\citenamefont {Glazov}\ and\ \citenamefont
  {Ganichev}(2014)}]{Glazov2014}%
  \BibitemOpen
  \bibfield  {author} {\bibinfo {author} {\bibfnamefont {M.}~\bibnamefont
  {Glazov}}\ and\ \bibinfo {author} {\bibfnamefont {S.}~\bibnamefont
  {Ganichev}},\ }\bibfield  {title} {\enquote {\bibinfo {title} {High frequency
  electric field induced nonlinear effects in graphene},}\ }\href {\doibase
  http://dx.doi.org/10.1016/j.physrep.2013.10.003} {\bibfield  {journal}
  {\bibinfo  {journal} {Physics Reports}\ }\textbf {\bibinfo {volume} {535}},\
  \bibinfo {pages} {101 } (\bibinfo {year} {2014})}\BibitemShut {NoStop}%
\bibitem [{\citenamefont {Hartmann}\ \emph {et~al.}(2014)\citenamefont
  {Hartmann}, \citenamefont {Kono},\ and\ \citenamefont
  {Portnoi}}]{Hartmann2013}%
  \BibitemOpen
  \bibfield  {author} {\bibinfo {author} {\bibfnamefont {R.~R.}\ \bibnamefont
  {Hartmann}}, \bibinfo {author} {\bibfnamefont {J.}~\bibnamefont {Kono}}, \
  and\ \bibinfo {author} {\bibfnamefont {M.~E.}\ \bibnamefont {Portnoi}},\
  }\bibfield  {title} {\enquote {\bibinfo {title} {Terahertz science and
  technology of carbon nanomaterials},}\ }\href {\doibase
  10.1088/0957-4484/25/32/322001} {\bibfield  {journal} {\bibinfo  {journal}
  {Nanotechnology}\ }\textbf {\bibinfo {volume} {25}},\ \bibinfo {pages}
  {322001} (\bibinfo {year} {2014})}\BibitemShut {NoStop}%
\bibitem [{\citenamefont {Oka}\ and\ \citenamefont {Aoki}(2009)}]{Oka2009}%
  \BibitemOpen
  \bibfield  {author} {\bibinfo {author} {\bibfnamefont {T.}~\bibnamefont
  {Oka}}\ and\ \bibinfo {author} {\bibfnamefont {H.}~\bibnamefont {Aoki}},\
  }\bibfield  {title} {\enquote {\bibinfo {title} {Photovoltaic hall effect in
  graphene},}\ }\href {http://link.aps.org/doi/10.1103/PhysRevB.79.081406}
  {\bibfield  {journal} {\bibinfo  {journal} {Phys. Rev. B}\ }\textbf {\bibinfo
  {volume} {79}},\ \bibinfo {pages} {081406} (\bibinfo {year}
  {2009})}\BibitemShut {NoStop}%
\bibitem [{\citenamefont {Calvo}\ \emph {et~al.}(2011)\citenamefont {Calvo},
  \citenamefont {Pastawski}, \citenamefont {Roche},\ and\ \citenamefont
  {Foa~Torres}}]{Calvo2011}%
  \BibitemOpen
  \bibfield  {author} {\bibinfo {author} {\bibfnamefont {H.~L.}\ \bibnamefont
  {Calvo}}, \bibinfo {author} {\bibfnamefont {H.~M.}\ \bibnamefont
  {Pastawski}}, \bibinfo {author} {\bibfnamefont {S.}~\bibnamefont {Roche}}, \
  and\ \bibinfo {author} {\bibfnamefont {L.~E.~F.}\ \bibnamefont
  {Foa~Torres}},\ }\bibfield  {title} {\enquote {\bibinfo {title} {Tuning
  laser-induced band gaps in graphene},}\ }\href
  {http://dx.doi.org/10.1063/1.3597412} {\bibfield  {journal} {\bibinfo
  {journal} {Appl. Phys. Lett.}\ }\textbf {\bibinfo {volume} {98}},\ \bibinfo
  {pages} {232103} (\bibinfo {year} {2011})}\BibitemShut {NoStop}%
\bibitem [{\citenamefont {Zhou}\ and\ \citenamefont {Wu}(2011)}]{Zhou2011}%
  \BibitemOpen
  \bibfield  {author} {\bibinfo {author} {\bibfnamefont {Y.}~\bibnamefont
  {Zhou}}\ and\ \bibinfo {author} {\bibfnamefont {M.~W.}\ \bibnamefont {Wu}},\
  }\bibfield  {title} {\enquote {\bibinfo {title} {Optical response of graphene
  under intense terahertz fields},}\ }\href
  {http://link.aps.org/doi/10.1103/PhysRevB.83.245436} {\bibfield  {journal}
  {\bibinfo  {journal} {Phys. Rev. B}\ }\textbf {\bibinfo {volume} {83}},\
  \bibinfo {pages} {245436} (\bibinfo {year} {2011})}\BibitemShut {NoStop}%
\bibitem [{\citenamefont {Savel’ev}\ and\ \citenamefont
  {Alexandrov}(2011)}]{Savelev2011}%
  \BibitemOpen
  \bibfield  {author} {\bibinfo {author} {\bibfnamefont {S.~E.}\ \bibnamefont
  {Savel’ev}}\ and\ \bibinfo {author} {\bibfnamefont {A.~S.}\ \bibnamefont
  {Alexandrov}},\ }\bibfield  {title} {\enquote {\bibinfo {title} {Massless
  dirac fermions in a laser field as a counterpart of graphene
  superlattices},}\ }\href {http://link.aps.org/doi/10.1103/PhysRevB.84.035428}
  {\bibfield  {journal} {\bibinfo  {journal} {Phys. Rev. B}\ }\textbf {\bibinfo
  {volume} {84}},\ \bibinfo {pages} {035428} (\bibinfo {year}
  {2011})}\BibitemShut {NoStop}%
\bibitem [{\citenamefont {Suarez~Morell}\ and\ \citenamefont
  {Foa~Torres}(2012)}]{SuarezMorell2012}%
  \BibitemOpen
  \bibfield  {author} {\bibinfo {author} {\bibfnamefont {E.}~\bibnamefont
  {Suarez~Morell}}\ and\ \bibinfo {author} {\bibfnamefont {L.~E.~F.}\
  \bibnamefont {Foa~Torres}},\ }\bibfield  {title} {\enquote {\bibinfo {title}
  {Radiation effects on the electric properties of bilayer graphene},}\
  }\href@noop {} {\bibfield  {journal} {\bibinfo  {journal} {Phys. Rev. B}\
  }\textbf {\bibinfo {volume} {86}},\ \bibinfo {pages} {125449} (\bibinfo
  {year} {2012})}\BibitemShut {NoStop}%
\bibitem [{\citenamefont {Wang}\ \emph {et~al.}(2013)\citenamefont {Wang},
  \citenamefont {Steinberg}, \citenamefont {Jarillo-Herrero},\ and\
  \citenamefont {Gedik}}]{Wang2013}%
  \BibitemOpen
  \bibfield  {author} {\bibinfo {author} {\bibfnamefont {Y.~H.}\ \bibnamefont
  {Wang}}, \bibinfo {author} {\bibfnamefont {H.}~\bibnamefont {Steinberg}},
  \bibinfo {author} {\bibfnamefont {P.}~\bibnamefont {Jarillo-Herrero}}, \ and\
  \bibinfo {author} {\bibfnamefont {N.}~\bibnamefont {Gedik}},\ }\bibfield
  {title} {\enquote {\bibinfo {title} {Observation of floquet-bloch states on
  the surface of a topological insulator},}\ }\href
  {http://www.sciencemag.org/content/342/6157/453.abstract} {\bibfield
  {journal} {\bibinfo  {journal} {Science}\ }\textbf {\bibinfo {volume}
  {342}},\ \bibinfo {pages} {453} (\bibinfo {year} {2013})}\BibitemShut
  {NoStop}%
\bibitem [{\citenamefont {Lindner}\ \emph {et~al.}(2011)\citenamefont
  {Lindner}, \citenamefont {Refael},\ and\ \citenamefont
  {Galitski}}]{Lindner2011}%
  \BibitemOpen
  \bibfield  {author} {\bibinfo {author} {\bibfnamefont {N.~H.}\ \bibnamefont
  {Lindner}}, \bibinfo {author} {\bibfnamefont {G.}~\bibnamefont {Refael}}, \
  and\ \bibinfo {author} {\bibfnamefont {V.}~\bibnamefont {Galitski}},\
  }\bibfield  {title} {\enquote {\bibinfo {title} {Floquet topological
  insulator in semiconductor quantum wells},}\ }\href
  {http://dx.doi.org/10.1038/nphys1926} {\bibfield  {journal} {\bibinfo
  {journal} {Nat Phys}\ }\textbf {\bibinfo {volume} {7}},\ \bibinfo {pages}
  {490} (\bibinfo {year} {2011})}\BibitemShut {NoStop}%
\bibitem [{\citenamefont {Kitagawa}\ \emph {et~al.}(2011)\citenamefont
  {Kitagawa}, \citenamefont {Oka}, \citenamefont {Brataas}, \citenamefont
  {Fu},\ and\ \citenamefont {Demler}}]{Kitagawa2011}%
  \BibitemOpen
  \bibfield  {author} {\bibinfo {author} {\bibfnamefont {T.}~\bibnamefont
  {Kitagawa}}, \bibinfo {author} {\bibfnamefont {T.}~\bibnamefont {Oka}},
  \bibinfo {author} {\bibfnamefont {A.}~\bibnamefont {Brataas}}, \bibinfo
  {author} {\bibfnamefont {L.}~\bibnamefont {Fu}}, \ and\ \bibinfo {author}
  {\bibfnamefont {E.}~\bibnamefont {Demler}},\ }\bibfield  {title} {\enquote
  {\bibinfo {title} {Transport properties of nonequilibrium systems under the
  application of light: Photoinduced quantum hall insulators without landau
  levels},}\ }\href {http://link.aps.org/doi/10.1103/PhysRevB.84.235108}
  {\bibfield  {journal} {\bibinfo  {journal} {Phys. Rev. B}\ }\textbf {\bibinfo
  {volume} {84}},\ \bibinfo {pages} {235108} (\bibinfo {year}
  {2011})}\BibitemShut {NoStop}%
\bibitem [{\citenamefont {Perez-Piskunow}\ \emph {et~al.}(2014)\citenamefont
  {Perez-Piskunow}, \citenamefont {Usaj}, \citenamefont {Balseiro},\ and\
  \citenamefont {Foa~Torres}}]{Piskunow2014}%
  \BibitemOpen
  \bibfield  {author} {\bibinfo {author} {\bibfnamefont {P.~M.}\ \bibnamefont
  {Perez-Piskunow}}, \bibinfo {author} {\bibfnamefont {G.}~\bibnamefont
  {Usaj}}, \bibinfo {author} {\bibfnamefont {C.~A.}\ \bibnamefont {Balseiro}},
  \ and\ \bibinfo {author} {\bibfnamefont {L.~E.~F.}\ \bibnamefont
  {Foa~Torres}},\ }\bibfield  {title} {\enquote {\bibinfo {title} {Floquet
  chiral edge states in graphene},}\ }\href {\doibase
  10.1103/PhysRevB.89.121401} {\bibfield  {journal} {\bibinfo  {journal} {Phys.
  Rev. B}\ }\textbf {\bibinfo {volume} {89}},\ \bibinfo {pages} {121401}
  (\bibinfo {year} {2014})}\BibitemShut {NoStop}%
\bibitem [{\citenamefont {Sentef}\ \emph {et~al.}(2015)\citenamefont {Sentef},
  \citenamefont {Claassen}, \citenamefont {Kemper}, \citenamefont {Moritz},
  \citenamefont {Oka}, \citenamefont {Freericks},\ and\ \citenamefont
  {Devereaux}}]{Sentef2014}%
  \BibitemOpen
  \bibfield  {author} {\bibinfo {author} {\bibfnamefont {M.~A.}\ \bibnamefont
  {Sentef}}, \bibinfo {author} {\bibfnamefont {M.}~\bibnamefont {Claassen}},
  \bibinfo {author} {\bibfnamefont {A.~F.}\ \bibnamefont {Kemper}}, \bibinfo
  {author} {\bibfnamefont {B.}~\bibnamefont {Moritz}}, \bibinfo {author}
  {\bibfnamefont {T.}~\bibnamefont {Oka}}, \bibinfo {author} {\bibfnamefont
  {J.~K.}\ \bibnamefont {Freericks}}, \ and\ \bibinfo {author} {\bibfnamefont
  {T.~P.}\ \bibnamefont {Devereaux}},\ }\bibfield  {title} {\enquote {\bibinfo
  {title} {Theory of floquet band formation and local pseudospin textures in
  pump-probe photoemission of graphene},}\ }\href@noop {} {\bibfield  {journal}
  {\bibinfo  {journal} {Nat Comm}\ }\textbf {\bibinfo {volume} {6}},\ \bibinfo
  {pages} {7047} (\bibinfo {year} {2015})}\BibitemShut {NoStop}%
\bibitem [{\citenamefont {Dahlhaus}\ \emph {et~al.}(2014)\citenamefont
  {Dahlhaus}, \citenamefont {Fregoso},\ and\ \citenamefont
  {Moore}}]{Dahlhaus2014}%
  \BibitemOpen
  \bibfield  {author} {\bibinfo {author} {\bibfnamefont {J.~P.}\ \bibnamefont
  {Dahlhaus}}, \bibinfo {author} {\bibfnamefont {B.~M.}\ \bibnamefont
  {Fregoso}}, \ and\ \bibinfo {author} {\bibfnamefont {J.~E.}\ \bibnamefont
  {Moore}},\ }\bibfield  {title} {\enquote {\bibinfo {title} {Magnetization
  signatures of light-induced quantum hall edge states},}\ }\href@noop {}
  {\bibfield  {journal} {\bibinfo  {journal} {arXiv:1408.6811}\ } (\bibinfo
  {year} {2014})}\BibitemShut {NoStop}%
\bibitem [{\citenamefont {Quelle}\ \emph {et~al.}(2014)\citenamefont {Quelle},
  \citenamefont {Beugeling},\ and\ \citenamefont {Smith}}]{Quelle2014}%
  \BibitemOpen
  \bibfield  {author} {\bibinfo {author} {\bibfnamefont {A.}~\bibnamefont
  {Quelle}}, \bibinfo {author} {\bibfnamefont {W.}~\bibnamefont {Beugeling}}, \
  and\ \bibinfo {author} {\bibfnamefont {C.~M.}\ \bibnamefont {Smith}},\
  }\bibfield  {title} {\enquote {\bibinfo {title} {Topological floquet states
  on a m\"obius band irradiated by circularly polarised light},}\ }\href@noop
  {} {\bibfield  {journal} {\bibinfo  {journal} {arXiv:1408.3087}\ } (\bibinfo
  {year} {2014})}\BibitemShut {NoStop}%
\bibitem [{\citenamefont {L\'opez}\ \emph {et~al.}(2015)\citenamefont
  {L\'opez}, \citenamefont {Scholz}, \citenamefont {Santos},\ and\
  \citenamefont {Schliemann}}]{Lopez2015}%
  \BibitemOpen
  \bibfield  {author} {\bibinfo {author} {\bibfnamefont {A.}~\bibnamefont
  {L\'opez}}, \bibinfo {author} {\bibfnamefont {A.}~\bibnamefont {Scholz}},
  \bibinfo {author} {\bibfnamefont {B.}~\bibnamefont {Santos}}, \ and\ \bibinfo
  {author} {\bibfnamefont {J.}~\bibnamefont {Schliemann}},\ }\bibfield  {title}
  {\enquote {\bibinfo {title} {Photoinduced pseudospin effects in silicene
  beyond the off-resonant condition},}\ }\href {\doibase
  10.1103/PhysRevB.91.125105} {\bibfield  {journal} {\bibinfo  {journal} {Phys.
  Rev. B}\ }\textbf {\bibinfo {volume} {91}},\ \bibinfo {pages} {125105}
  (\bibinfo {year} {2015})}\BibitemShut {NoStop}%
\bibitem [{\citenamefont {Goldman}\ and\ \citenamefont
  {Dalibard}(2014)}]{Goldman2014}%
  \BibitemOpen
  \bibfield  {author} {\bibinfo {author} {\bibfnamefont {N.}~\bibnamefont
  {Goldman}}\ and\ \bibinfo {author} {\bibfnamefont {J.}~\bibnamefont
  {Dalibard}},\ }\bibfield  {title} {\enquote {\bibinfo {title} {Periodically
  driven quantum systems: Effective hamiltonians and engineered gauge
  fields},}\ }\href {http://link.aps.org/doi/10.1103/PhysRevX.4.031027}
  {\bibfield  {journal} {\bibinfo  {journal} {Phys. Rev. X}\ }\textbf {\bibinfo
  {volume} {4}},\ \bibinfo {pages} {031027} (\bibinfo {year}
  {2014})}\BibitemShut {NoStop}%
\bibitem [{\citenamefont {Choudhury}\ and\ \citenamefont
  {Mueller}(2014)}]{Choudhury2014}%
  \BibitemOpen
  \bibfield  {author} {\bibinfo {author} {\bibfnamefont {S.}~\bibnamefont
  {Choudhury}}\ and\ \bibinfo {author} {\bibfnamefont {E.~J.}\ \bibnamefont
  {Mueller}},\ }\bibfield  {title} {\enquote {\bibinfo {title} {Stability of a
  floquet bose-einstein condensate in a one-dimensional optical lattice},}\
  }\href {\doibase http://dx.doi.org/10.1103/PhysRevA.90.013621} {\bibfield
  {journal} {\bibinfo  {journal} {Phys. Rev. A}\ }\textbf {\bibinfo {volume}
  {90}},\ \bibinfo {pages} {013621} (\bibinfo {year} {2014})}\BibitemShut
  {NoStop}%
\bibitem [{\citenamefont {G\'omez-Le\'on}\ \emph {et~al.}(2014)\citenamefont
  {G\'omez-Le\'on}, \citenamefont {Delplace},\ and\ \citenamefont
  {Platero}}]{Gomez-Leon2014}%
  \BibitemOpen
  \bibfield  {author} {\bibinfo {author} {\bibfnamefont {A.}~\bibnamefont
  {G\'omez-Le\'on}}, \bibinfo {author} {\bibfnamefont {P.}~\bibnamefont
  {Delplace}}, \ and\ \bibinfo {author} {\bibfnamefont {G.}~\bibnamefont
  {Platero}},\ }\bibfield  {title} {\enquote {\bibinfo {title} {Engineering
  anomalous quantum hall plateaus and antichiral states with ac fields},}\
  }\href {http://link.aps.org/doi/10.1103/PhysRevB.89.205408} {\bibfield
  {journal} {\bibinfo  {journal} {Phys. Rev. B}\ }\textbf {\bibinfo {volume}
  {89}},\ \bibinfo {pages} {205408} (\bibinfo {year} {2014})}\BibitemShut
  {NoStop}%
\bibitem [{\citenamefont {Bilitewski}\ and\ \citenamefont
  {Cooper}(2015)}]{Bilitewski2015}%
  \BibitemOpen
  \bibfield  {author} {\bibinfo {author} {\bibfnamefont {T.}~\bibnamefont
  {Bilitewski}}\ and\ \bibinfo {author} {\bibfnamefont {N.~R.}\ \bibnamefont
  {Cooper}},\ }\bibfield  {title} {\enquote {\bibinfo {title} {Scattering
  theory for floquet-bloch states},}\ }\href {\doibase
  10.1103/PhysRevA.91.033601} {\bibfield  {journal} {\bibinfo  {journal} {Phys.
  Rev. A}\ }\textbf {\bibinfo {volume} {91}},\ \bibinfo {pages} {033601}
  (\bibinfo {year} {2015})}\BibitemShut {NoStop}%
\bibitem [{\citenamefont {Tenenbaum~Katan}\ and\ \citenamefont
  {Podolsky}(2013{\natexlab{a}})}]{TenenbaumKatan2013a}%
  \BibitemOpen
  \bibfield  {author} {\bibinfo {author} {\bibfnamefont {Y.}~\bibnamefont
  {Tenenbaum~Katan}}\ and\ \bibinfo {author} {\bibfnamefont {D.}~\bibnamefont
  {Podolsky}},\ }\bibfield  {title} {\enquote {\bibinfo {title} {Modulated
  floquet topological insulators},}\ }\href
  {http://link.aps.org/doi/10.1103/PhysRevLett.110.016802} {\bibfield
  {journal} {\bibinfo  {journal} {Phys. Rev. Lett.}\ }\textbf {\bibinfo
  {volume} {110}},\ \bibinfo {pages} {016802} (\bibinfo {year}
  {2013}{\natexlab{a}})}\BibitemShut {NoStop}%
\bibitem [{\citenamefont {Rudner}\ \emph {et~al.}(2013)\citenamefont {Rudner},
  \citenamefont {Lindner}, \citenamefont {Berg},\ and\ \citenamefont
  {Levin}}]{Rudner2013}%
  \BibitemOpen
  \bibfield  {author} {\bibinfo {author} {\bibfnamefont {M.~S.}\ \bibnamefont
  {Rudner}}, \bibinfo {author} {\bibfnamefont {N.~H.}\ \bibnamefont {Lindner}},
  \bibinfo {author} {\bibfnamefont {E.}~\bibnamefont {Berg}}, \ and\ \bibinfo
  {author} {\bibfnamefont {M.}~\bibnamefont {Levin}},\ }\bibfield  {title}
  {\enquote {\bibinfo {title} {Anomalous edge states and the bulk-edge
  correspondence for periodically-driven two dimensional systems},}\
  }\href@noop {} {\bibfield  {journal} {\bibinfo  {journal} {Phys. Rev. X}\
  }\textbf {\bibinfo {volume} {3}},\ \bibinfo {pages} {031005} (\bibinfo {year}
  {2013})}\BibitemShut {NoStop}%
\bibitem [{\citenamefont {Ho}\ and\ \citenamefont {Gong}(2014)}]{Ho2014}%
  \BibitemOpen
  \bibfield  {author} {\bibinfo {author} {\bibfnamefont {D.~Y.~H.}\
  \bibnamefont {Ho}}\ and\ \bibinfo {author} {\bibfnamefont {J.}~\bibnamefont
  {Gong}},\ }\bibfield  {title} {\enquote {\bibinfo {title} {Topological
  effects in chiral symmetric driven systems},}\ }\href {\doibase
  10.1103/PhysRevB.90.195419} {\bibfield  {journal} {\bibinfo  {journal} {Phys.
  Rev. B}\ }\textbf {\bibinfo {volume} {90}},\ \bibinfo {pages} {195419}
  (\bibinfo {year} {2014})}\BibitemShut {NoStop}%
\bibitem [{\citenamefont {Zhou}\ \emph {et~al.}(2014)\citenamefont {Zhou},
  \citenamefont {Wang}, \citenamefont {Ho},\ and\ \citenamefont
  {Gong}}]{Zhou2014}%
  \BibitemOpen
  \bibfield  {author} {\bibinfo {author} {\bibfnamefont {L.}~\bibnamefont
  {Zhou}}, \bibinfo {author} {\bibfnamefont {H.}~\bibnamefont {Wang}}, \bibinfo
  {author} {\bibfnamefont {D.~Y.}\ \bibnamefont {Ho}}, \ and\ \bibinfo {author}
  {\bibfnamefont {J.}~\bibnamefont {Gong}},\ }\bibfield  {title} {\enquote
  {\bibinfo {title} {Aspects of floquet bands and topological phase transitions
  in a continuously driven superlattice},}\ }\href
  {http://dx.doi.org/10.1140/epjb/e2014-50465-9} {\bibfield  {journal}
  {\bibinfo  {journal} {Eur. Phys. J. B}\ }\textbf {\bibinfo {volume} {87}},\
  \bibinfo {pages} {204} (\bibinfo {year} {2014})}\BibitemShut {NoStop}%
\bibitem [{\citenamefont {Usaj}\ \emph {et~al.}(2014)\citenamefont {Usaj},
  \citenamefont {Perez-Piskunow}, \citenamefont {Foa~Torres},\ and\
  \citenamefont {Balseiro}}]{Usaj2014}%
  \BibitemOpen
  \bibfield  {author} {\bibinfo {author} {\bibfnamefont {G.}~\bibnamefont
  {Usaj}}, \bibinfo {author} {\bibfnamefont {P.~M.}\ \bibnamefont
  {Perez-Piskunow}}, \bibinfo {author} {\bibfnamefont {L.~E.~F.}\ \bibnamefont
  {Foa~Torres}}, \ and\ \bibinfo {author} {\bibfnamefont {C.~A.}\ \bibnamefont
  {Balseiro}},\ }\bibfield  {title} {\enquote {\bibinfo {title} {Irradiated
  graphene as a tunable floquet topological insulator},}\ }\href {\doibase
  10.1103/PhysRevB.90.115423} {\bibfield  {journal} {\bibinfo  {journal} {Phys.
  Rev. B}\ }\textbf {\bibinfo {volume} {90}},\ \bibinfo {pages} {115423}
  (\bibinfo {year} {2014})}\BibitemShut {NoStop}%
\bibitem [{\citenamefont {D'Alessio}\ and\ \citenamefont
  {Rigol}(2014)}]{DAlessio2014}%
  \BibitemOpen
  \bibfield  {author} {\bibinfo {author} {\bibfnamefont {L.}~\bibnamefont
  {D'Alessio}}\ and\ \bibinfo {author} {\bibfnamefont {M.}~\bibnamefont
  {Rigol}},\ }\bibfield  {title} {\enquote {\bibinfo {title} {Dynamical
  preparation of floquet chern insulators:a no-go theorem, the bott index, and
  boundary effects},}\ }\href@noop {} {\bibfield  {journal} {\bibinfo
  {journal} {arXiv:1409.6319}\ } (\bibinfo {year} {2014})}\BibitemShut
  {NoStop}%
\bibitem [{\citenamefont {Dehghani}\ \emph {et~al.}(2014)\citenamefont
  {Dehghani}, \citenamefont {Oka},\ and\ \citenamefont {Mitra}}]{Dehghani2014}%
  \BibitemOpen
  \bibfield  {author} {\bibinfo {author} {\bibfnamefont {H.}~\bibnamefont
  {Dehghani}}, \bibinfo {author} {\bibfnamefont {T.}~\bibnamefont {Oka}}, \
  and\ \bibinfo {author} {\bibfnamefont {A.}~\bibnamefont {Mitra}},\ }\bibfield
   {title} {\enquote {\bibinfo {title} {Dissipative floquet topological
  systems},}\ }\href {http://link.aps.org/doi/10.1103/PhysRevB.90.195429}
  {\bibfield  {journal} {\bibinfo  {journal} {Phys. Rev. B}\ }\textbf {\bibinfo
  {volume} {90}},\ \bibinfo {pages} {195429} (\bibinfo {year}
  {2014})}\BibitemShut {NoStop}%
\bibitem [{\citenamefont {Liu}(2015)}]{Liu2015}%
  \BibitemOpen
  \bibfield  {author} {\bibinfo {author} {\bibfnamefont {D.~E.}\ \bibnamefont
  {Liu}},\ }\bibfield  {title} {\enquote {\bibinfo {title} {Classification of
  the floquet statistical distribution for time-periodic open systems},}\
  }\href {\doibase 10.1103/PhysRevB.91.144301} {\bibfield  {journal} {\bibinfo
  {journal} {Phys. Rev. B}\ }\textbf {\bibinfo {volume} {91}},\ \bibinfo
  {pages} {144301} (\bibinfo {year} {2015})}\BibitemShut {NoStop}%
\bibitem [{\citenamefont {Seetharam}\ \emph {et~al.}(2015)\citenamefont
  {Seetharam}, \citenamefont {Bardyn}, \citenamefont {Lindner}, \citenamefont
  {Rudner},\ and\ \citenamefont {Refael}}]{Seetharam2015}%
  \BibitemOpen
  \bibfield  {author} {\bibinfo {author} {\bibfnamefont {K.~I.}\ \bibnamefont
  {Seetharam}}, \bibinfo {author} {\bibfnamefont {C.-E.}\ \bibnamefont
  {Bardyn}}, \bibinfo {author} {\bibfnamefont {N.~H.}\ \bibnamefont {Lindner}},
  \bibinfo {author} {\bibfnamefont {M.~S.}\ \bibnamefont {Rudner}}, \ and\
  \bibinfo {author} {\bibfnamefont {G.}~\bibnamefont {Refael}},\ }\bibfield
  {title} {\enquote {\bibinfo {title} {Controlled population of floquet-bloch
  states via coupling to bose and fermi baths},}\ }\href@noop {} {\bibfield
  {journal} {\bibinfo  {journal} {arXiv:1502.02664}\ } (\bibinfo {year}
  {2015})}\BibitemShut {NoStop}%
\bibitem [{\citenamefont {Iadecola}\ \emph {et~al.}(2015)\citenamefont
  {Iadecola}, \citenamefont {Neupert},\ and\ \citenamefont
  {Chamon}}]{Iadecola2015}%
  \BibitemOpen
  \bibfield  {author} {\bibinfo {author} {\bibfnamefont {T.}~\bibnamefont
  {Iadecola}}, \bibinfo {author} {\bibfnamefont {T.}~\bibnamefont {Neupert}}, \
  and\ \bibinfo {author} {\bibfnamefont {C.}~\bibnamefont {Chamon}},\
  }\bibfield  {title} {\enquote {\bibinfo {title} {Occupation of topological
  floquet bands in open systems},}\ }\href@noop {} {\bibfield  {journal}
  {\bibinfo  {journal} {arXiv:1502.05047}\ } (\bibinfo {year}
  {2015})}\BibitemShut {NoStop}%
\bibitem [{\citenamefont {Gu}\ \emph {et~al.}(2011)\citenamefont {Gu},
  \citenamefont {Fertig}, \citenamefont {Arovas},\ and\ \citenamefont
  {Auerbach}}]{Gu2011}%
  \BibitemOpen
  \bibfield  {author} {\bibinfo {author} {\bibfnamefont {Z.}~\bibnamefont
  {Gu}}, \bibinfo {author} {\bibfnamefont {H.~A.}\ \bibnamefont {Fertig}},
  \bibinfo {author} {\bibfnamefont {D.~P.}\ \bibnamefont {Arovas}}, \ and\
  \bibinfo {author} {\bibfnamefont {A.}~\bibnamefont {Auerbach}},\ }\bibfield
  {title} {\enquote {\bibinfo {title} {Floquet spectrum and transport through
  an irradiated graphene ribbon},}\ }\href
  {http://link.aps.org/doi/10.1103/PhysRevLett.107.216601} {\bibfield
  {journal} {\bibinfo  {journal} {Phys. Rev. Lett.}\ }\textbf {\bibinfo
  {volume} {107}},\ \bibinfo {pages} {216601} (\bibinfo {year}
  {2011})}\BibitemShut {NoStop}%
\bibitem [{\citenamefont {Kundu}\ \emph {et~al.}(2014)\citenamefont {Kundu},
  \citenamefont {Fertig},\ and\ \citenamefont {Seradjeh}}]{Kundu2014}%
  \BibitemOpen
  \bibfield  {author} {\bibinfo {author} {\bibfnamefont {A.}~\bibnamefont
  {Kundu}}, \bibinfo {author} {\bibfnamefont {H.~A.}\ \bibnamefont {Fertig}}, \
  and\ \bibinfo {author} {\bibfnamefont {B.}~\bibnamefont {Seradjeh}},\
  }\bibfield  {title} {\enquote {\bibinfo {title} {Effective theory of floquet
  topological transitions},}\ }\href
  {http://link.aps.org/doi/10.1103/PhysRevLett.113.236803} {\bibfield
  {journal} {\bibinfo  {journal} {Phys. Rev. Lett.}\ }\textbf {\bibinfo
  {volume} {113}},\ \bibinfo {pages} {236803} (\bibinfo {year}
  {2014})}\BibitemShut {NoStop}%
\bibitem [{\citenamefont {Foa~Torres}\ \emph {et~al.}(2014)\citenamefont
  {Foa~Torres}, \citenamefont {Perez-Piskunow}, \citenamefont {Balseiro},\ and\
  \citenamefont {Usaj}}]{Foa2014}%
  \BibitemOpen
  \bibfield  {author} {\bibinfo {author} {\bibfnamefont {L.~E.~F.}\
  \bibnamefont {Foa~Torres}}, \bibinfo {author} {\bibfnamefont {P.~M.}\
  \bibnamefont {Perez-Piskunow}}, \bibinfo {author} {\bibfnamefont {C.~A.}\
  \bibnamefont {Balseiro}}, \ and\ \bibinfo {author} {\bibfnamefont
  {G.}~\bibnamefont {Usaj}},\ }\bibfield  {title} {\enquote {\bibinfo {title}
  {Multiterminal conductance of a floquet topological insulator},}\ }\href
  {http://journals.aps.org/prl/abstract/10.1103/PhysRevLett.113.266801}
  {\bibfield  {journal} {\bibinfo  {journal} {Phys. Rev. Lett.}\ }\textbf
  {\bibinfo {volume} {113}},\ \bibinfo {pages} {266801} (\bibinfo {year}
  {2014})}\BibitemShut {NoStop}%
\bibitem [{\citenamefont {Dehghani}\ \emph {et~al.}(2015)\citenamefont
  {Dehghani}, \citenamefont {Oka},\ and\ \citenamefont {Mitra}}]{Dehghani2015}%
  \BibitemOpen
  \bibfield  {author} {\bibinfo {author} {\bibfnamefont {H.}~\bibnamefont
  {Dehghani}}, \bibinfo {author} {\bibfnamefont {T.}~\bibnamefont {Oka}}, \
  and\ \bibinfo {author} {\bibfnamefont {A.}~\bibnamefont {Mitra}},\ }\bibfield
   {title} {\enquote {\bibinfo {title} {Out-of-equilibrium electrons and the
  hall conductance of a floquet topological insulator},}\ }\href {\doibase
  10.1103/PhysRevB.91.155422} {\bibfield  {journal} {\bibinfo  {journal} {Phys.
  Rev. B}\ }\textbf {\bibinfo {volume} {91}},\ \bibinfo {pages} {155422}
  (\bibinfo {year} {2015})}\BibitemShut {NoStop}%
\bibitem [{\citenamefont {Lindner}\ \emph {et~al.}(2013)\citenamefont
  {Lindner}, \citenamefont {Bergman}, \citenamefont {Refael},\ and\
  \citenamefont {Galitski}}]{Lindner2013}%
  \BibitemOpen
  \bibfield  {author} {\bibinfo {author} {\bibfnamefont {N.~H.}\ \bibnamefont
  {Lindner}}, \bibinfo {author} {\bibfnamefont {D.~L.}\ \bibnamefont
  {Bergman}}, \bibinfo {author} {\bibfnamefont {G.}~\bibnamefont {Refael}}, \
  and\ \bibinfo {author} {\bibfnamefont {V.}~\bibnamefont {Galitski}},\
  }\bibfield  {title} {\enquote {\bibinfo {title} {Topological floquet spectrum
  in three dimensions via a two-photon resonance},}\ }\href
  {http://link.aps.org/doi/10.1103/PhysRevB.87.235131} {\bibfield  {journal}
  {\bibinfo  {journal} {Phys. Rev. B}\ }\textbf {\bibinfo {volume} {87}},\
  \bibinfo {pages} {235131} (\bibinfo {year} {2013})}\BibitemShut {NoStop}%
\bibitem [{\citenamefont {Tenenbaum~Katan}\ and\ \citenamefont
  {Podolsky}(2013{\natexlab{b}})}]{Tenenbaum-Katan2013}%
  \BibitemOpen
  \bibfield  {author} {\bibinfo {author} {\bibfnamefont {Y.}~\bibnamefont
  {Tenenbaum~Katan}}\ and\ \bibinfo {author} {\bibfnamefont {D.}~\bibnamefont
  {Podolsky}},\ }\bibfield  {title} {\enquote {\bibinfo {title} {Generation and
  manipulation of localized modes in floquet topological insulators},}\ }\href
  {\doibase 10.1103/PhysRevB.88.224106} {\bibfield  {journal} {\bibinfo
  {journal} {Phys. Rev. B}\ }\textbf {\bibinfo {volume} {88}},\ \bibinfo
  {pages} {224106} (\bibinfo {year} {2013}{\natexlab{b}})}\BibitemShut
  {NoStop}%
\bibitem [{\citenamefont {Zhang}\ \emph {et~al.}(2012)\citenamefont {Zhang},
  \citenamefont {Kane},\ and\ \citenamefont {Mele}}]{Zhang2012}%
  \BibitemOpen
  \bibfield  {author} {\bibinfo {author} {\bibfnamefont {F.}~\bibnamefont
  {Zhang}}, \bibinfo {author} {\bibfnamefont {C.~L.}\ \bibnamefont {Kane}}, \
  and\ \bibinfo {author} {\bibfnamefont {E.~J.}\ \bibnamefont {Mele}},\
  }\bibfield  {title} {\enquote {\bibinfo {title} {Surface states of
  topological insulators},}\ }\href {\doibase 10.1103/PhysRevB.86.081303}
  {\bibfield  {journal} {\bibinfo  {journal} {Phys. Rev. B}\ }\textbf {\bibinfo
  {volume} {86}},\ \bibinfo {pages} {081303} (\bibinfo {year}
  {2012})}\BibitemShut {NoStop}%
\bibitem [{\citenamefont {Fregoso}\ \emph {et~al.}(2013)\citenamefont
  {Fregoso}, \citenamefont {Wang}, \citenamefont {Gedik},\ and\ \citenamefont
  {Galitski}}]{Fregoso2013}%
  \BibitemOpen
  \bibfield  {author} {\bibinfo {author} {\bibfnamefont {B.~M.}\ \bibnamefont
  {Fregoso}}, \bibinfo {author} {\bibfnamefont {Y.~H.}\ \bibnamefont {Wang}},
  \bibinfo {author} {\bibfnamefont {N.}~\bibnamefont {Gedik}}, \ and\ \bibinfo
  {author} {\bibfnamefont {V.}~\bibnamefont {Galitski}},\ }\bibfield  {title}
  {\enquote {\bibinfo {title} {Driven electronic states at the surface of a
  topological insulator},}\ }\href {\doibase 10.1103/PhysRevB.88.155129}
  {\bibfield  {journal} {\bibinfo  {journal} {Phys. Rev. B}\ }\textbf {\bibinfo
  {volume} {88}},\ \bibinfo {pages} {155129} (\bibinfo {year}
  {2013})}\BibitemShut {NoStop}%
\bibitem [{\citenamefont {Syzranov}\ \emph {et~al.}(2008)\citenamefont
  {Syzranov}, \citenamefont {Fistul},\ and\ \citenamefont
  {Efetov}}]{Syzranov2008}%
  \BibitemOpen
  \bibfield  {author} {\bibinfo {author} {\bibfnamefont {S.~V.}\ \bibnamefont
  {Syzranov}}, \bibinfo {author} {\bibfnamefont {M.~V.}\ \bibnamefont
  {Fistul}}, \ and\ \bibinfo {author} {\bibfnamefont {K.~B.}\ \bibnamefont
  {Efetov}},\ }\bibfield  {title} {\enquote {\bibinfo {title} {Effect of
  radiation on transport in graphene},}\ }\href
  {http://link.aps.org/doi/10.1103/PhysRevB.78.045407} {\bibfield  {journal}
  {\bibinfo  {journal} {Phys. Rev. B}\ }\textbf {\bibinfo {volume} {78}},\
  \bibinfo {pages} {045407} (\bibinfo {year} {2008})}\BibitemShut {NoStop}%
\bibitem [{\citenamefont {Fu}\ and\ \citenamefont {Kane}(2008)}]{Fu2008}%
  \BibitemOpen
  \bibfield  {author} {\bibinfo {author} {\bibfnamefont {L.}~\bibnamefont
  {Fu}}\ and\ \bibinfo {author} {\bibfnamefont {C.~L.}\ \bibnamefont {Kane}},\
  }\bibfield  {title} {\enquote {\bibinfo {title} {Superconducting proximity
  effect and majorana fermions at the surface of a topological insulator},}\
  }\href {\doibase 10.1103/PhysRevLett.100.096407} {\bibfield  {journal}
  {\bibinfo  {journal} {Phys. Rev. Lett.}\ }\textbf {\bibinfo {volume} {100}},\
  \bibinfo {pages} {096407} (\bibinfo {year} {2008})}\BibitemShut {NoStop}%
\bibitem [{\citenamefont {Chu}\ \emph {et~al.}(2011)\citenamefont {Chu},
  \citenamefont {Shi},\ and\ \citenamefont {Shen}}]{Chu2011}%
  \BibitemOpen
  \bibfield  {author} {\bibinfo {author} {\bibfnamefont {R.-L.}\ \bibnamefont
  {Chu}}, \bibinfo {author} {\bibfnamefont {J.}~\bibnamefont {Shi}}, \ and\
  \bibinfo {author} {\bibfnamefont {S.-Q.}\ \bibnamefont {Shen}},\ }\bibfield
  {title} {\enquote {\bibinfo {title} {Surface edge state and half-quantized
  hall conductance in topological insulators},}\ }\href {\doibase
  10.1103/PhysRevB.84.085312} {\bibfield  {journal} {\bibinfo  {journal} {Phys.
  Rev. B}\ }\textbf {\bibinfo {volume} {84}},\ \bibinfo {pages} {085312}
  (\bibinfo {year} {2011})}\BibitemShut {NoStop}%
\bibitem [{\citenamefont {Zhang}\ \emph {et~al.}(2013)\citenamefont {Zhang},
  \citenamefont {Kane},\ and\ \citenamefont {Mele}}]{Zhang2013}%
  \BibitemOpen
  \bibfield  {author} {\bibinfo {author} {\bibfnamefont {F.}~\bibnamefont
  {Zhang}}, \bibinfo {author} {\bibfnamefont {C.~L.}\ \bibnamefont {Kane}}, \
  and\ \bibinfo {author} {\bibfnamefont {E.~J.}\ \bibnamefont {Mele}},\
  }\bibfield  {title} {\enquote {\bibinfo {title} {Surface state magnetization
  and chiral edge states on topological insulators},}\ }\href {\doibase
  10.1103/PhysRevLett.110.046404} {\bibfield  {journal} {\bibinfo  {journal}
  {Phys. Rev. Lett.}\ }\textbf {\bibinfo {volume} {110}},\ \bibinfo {pages}
  {046404} (\bibinfo {year} {2013})}\BibitemShut {NoStop}%
\bibitem [{\citenamefont {Perez-Piskunow}\ \emph {et~al.}(2015)\citenamefont
  {Perez-Piskunow}, \citenamefont {Foa~Torres},\ and\ \citenamefont
  {Usaj}}]{Piskunow2015}%
  \BibitemOpen
  \bibfield  {author} {\bibinfo {author} {\bibfnamefont {P.~M.}\ \bibnamefont
  {Perez-Piskunow}}, \bibinfo {author} {\bibfnamefont {L.~E.~F.}\ \bibnamefont
  {Foa~Torres}}, \ and\ \bibinfo {author} {\bibfnamefont {G.}~\bibnamefont
  {Usaj}},\ }\bibfield  {title} {\enquote {\bibinfo {title} {Hierarchy of
  floquet gaps and edge states for driven honeycomb lattices},}\ }\href
  {\doibase 10.1103/PhysRevA.91.043625} {\bibfield  {journal} {\bibinfo
  {journal} {Phys. Rev. A}\ }\textbf {\bibinfo {volume} {91}},\ \bibinfo
  {pages} {043625} (\bibinfo {year} {2015})}\BibitemShut {NoStop}%
\bibitem [{\citenamefont {Niemi}\ and\ \citenamefont
  {Semenoff}(1983)}]{Niemi1983}%
  \BibitemOpen
  \bibfield  {author} {\bibinfo {author} {\bibfnamefont {A.~J.}\ \bibnamefont
  {Niemi}}\ and\ \bibinfo {author} {\bibfnamefont {G.~W.}\ \bibnamefont
  {Semenoff}},\ }\bibfield  {title} {\enquote {\bibinfo {title}
  {Axial-anomaly-induced fermion fractionization and effective gauge-theory
  actions in odd-dimensional space-times},}\ }\href {\doibase
  10.1103/PhysRevLett.51.2077} {\bibfield  {journal} {\bibinfo  {journal}
  {Phys. Rev. Lett.}\ }\textbf {\bibinfo {volume} {51}},\ \bibinfo {pages}
  {2077} (\bibinfo {year} {1983})}\BibitemShut {NoStop}%
\bibitem [{\citenamefont {Redlich}(1984)}]{Redlich1984}%
  \BibitemOpen
  \bibfield  {author} {\bibinfo {author} {\bibfnamefont {A.~N.}\ \bibnamefont
  {Redlich}},\ }\bibfield  {title} {\enquote {\bibinfo {title} {Parity
  violation and gauge noninvariance of the effective gauge field action in
  three dimensions},}\ }\href {\doibase 10.1103/PhysRevD.29.2366} {\bibfield
  {journal} {\bibinfo  {journal} {Phys. Rev. D}\ }\textbf {\bibinfo {volume}
  {29}},\ \bibinfo {pages} {2366} (\bibinfo {year} {1984})}\BibitemShut
  {NoStop}%
\bibitem [{\citenamefont {Qi}\ \emph {et~al.}(2008)\citenamefont {Qi},
  \citenamefont {Hughes},\ and\ \citenamefont {Zhang}}]{Qi2008}%
  \BibitemOpen
  \bibfield  {author} {\bibinfo {author} {\bibfnamefont {X.-L.}\ \bibnamefont
  {Qi}}, \bibinfo {author} {\bibfnamefont {T.~L.}\ \bibnamefont {Hughes}}, \
  and\ \bibinfo {author} {\bibfnamefont {S.-C.}\ \bibnamefont {Zhang}},\
  }\bibfield  {title} {\enquote {\bibinfo {title} {Topological field theory of
  time-reversal invariant insulators},}\ }\href {\doibase
  10.1103/PhysRevB.78.195424} {\bibfield  {journal} {\bibinfo  {journal} {Phys.
  Rev. B}\ }\textbf {\bibinfo {volume} {78}},\ \bibinfo {pages} {195424}
  (\bibinfo {year} {2008})}\BibitemShut {NoStop}%
\bibitem [{\citenamefont {Zhang}\ \emph {et~al.}(2015)\citenamefont {Zhang},
  \citenamefont {Lu},\ and\ \citenamefont {Shen}}]{Zhang2015}%
  \BibitemOpen
  \bibfield  {author} {\bibinfo {author} {\bibfnamefont {S.-B.}\ \bibnamefont
  {Zhang}}, \bibinfo {author} {\bibfnamefont {H.-Z.}\ \bibnamefont {Lu}}, \
  and\ \bibinfo {author} {\bibfnamefont {S.-Q.}\ \bibnamefont {Shen}},\
  }\bibfield  {title} {\enquote {\bibinfo {title} {Edge states and integer
  quantum hall effect in topological insulator thin films},}\ }\href@noop {}
  {\bibfield  {journal} {\bibinfo  {journal} {arXiv:1502.01792}\ } (\bibinfo
  {year} {2015})}\BibitemShut {NoStop}%
\bibitem [{not({\natexlab{a}})}]{note1}%
  \BibitemOpen
  \href@noop {} {} {Here we exemplify in the
  truncated basis $m \in \{0, 1 \}$ for the ZB gap region. For the ZC gap we
  include the $m = -1$ channel to keep the truncated basis symmetric with
  respect to the level crossing at $\varepsilon =
  0$~\cite{Piskunow2014,Usaj2014}. The general resolution of the differential
  equation is, however, completely analogous.}\BibitemShut {Stop}%
\bibitem [{not({\natexlab{b}})}]{note2}%
  \BibitemOpen
  \href@noop {} {} {See Supplemental
  Material at [URL will be inserted by publisher] for more
  details.}\BibitemShut {Stop}%
\bibitem [{\citenamefont {Zhang}\ \emph {et~al.}(2009)\citenamefont {Zhang},
  \citenamefont {Liu}, \citenamefont {Qi}, \citenamefont {Dai}, \citenamefont
  {Fang},\ and\ \citenamefont {Zhang}}]{Zhang2009}%
  \BibitemOpen
  \bibfield  {author} {\bibinfo {author} {\bibfnamefont {H.}~\bibnamefont
  {Zhang}}, \bibinfo {author} {\bibfnamefont {C.-X.}\ \bibnamefont {Liu}},
  \bibinfo {author} {\bibfnamefont {X.-L.}\ \bibnamefont {Qi}}, \bibinfo
  {author} {\bibfnamefont {X.}~\bibnamefont {Dai}}, \bibinfo {author}
  {\bibfnamefont {Z.}~\bibnamefont {Fang}}, \ and\ \bibinfo {author}
  {\bibfnamefont {S.-C.}\ \bibnamefont {Zhang}},\ }\bibfield  {title} {\enquote
  {\bibinfo {title} {Topological insulators in $\text{Bi}_2\text{Se}_3$,
  $\text{Bi}_2\text{Te}_3$ and $\text{Sb}_2\text{Te}_3$ with a single dirac
  cone on the surface},}\ }\href {\doibase 10.1038/nphys1270} {\bibfield
  {journal} {\bibinfo  {journal} {Nat. Phys.}\ }\textbf {\bibinfo {volume}
  {5}},\ \bibinfo {pages} {438} (\bibinfo {year} {2009})}\BibitemShut {NoStop}%
\bibitem [{not({\natexlab{c}})}]{note4}%
  \BibitemOpen
  \href@noop {} {} {This was numerically
  corroborated in a configuration with two laser beams with different
  frequencies and opposite helicities, see~\cite{note2}. Although the
  time-reversal symmetry is broken everywhere, when continuously connecting the
  two distinct topological phases the laser-induced band gap necessarily closes
  at some point in between the two regions~\cite{Lindner2011}.}\BibitemShut
  {Stop}%
\bibitem [{not({\natexlab{d}})}]{note3}%
  \BibitemOpen
  \href@noop {} {} {Although this condition
  seems particular to the specific Floquet Hamiltonian in
  Eq.~(\ref{eq:Floquet}), it is not necessarily required for the resolution of
  Eq.~(\ref{eq:diffeq}).}\BibitemShut {Stop}%
\bibitem [{not({\natexlab{e}})}]{note5}%
  \BibitemOpen
  \href@noop {} {} {This is also the case
  for the interface states crossing the Floquet zone boundary gap. In this
  situation, the difference in the frequencies needs to be smaller than the
  width of the gap.}\BibitemShut {Stop}%
\end{thebibliography}%

\end{document}